\documentclass[aps,
 prx,amsmath,amssymb,showpacs,showkeys,longbibliography,
 reprint,
]{revtex4-1}

\usepackage{graphicx}
\usepackage{dcolumn}
\usepackage{bm}
\usepackage{color}
\usepackage[svgnames,dvipsnames]{xcolor}   
\usepackage{ulem}
\usepackage{oplotsymbl}
\usepackage{upgreek}

\begin{document}


\title[Sample title]{A high-sensitivity resistance bridge for nanoscale thermal microscopy}


\author{Wilfrid Poirier$^{1}$, Victor Guillemot$^{2}$, Mohammed Mghalfi$^{1}$, Valentina Krachmalnicoff$^{2}$, Yannick De Wilde$^{2}$}
\address{$^{1}$ Laboratoire national de m\'etrologie et d'essais, 29 avenue Roger Hennequin, 78197 Trappes, France}
\address{$^{2}$ Institut Langevin, ESPCI Paris, Université PSL, CNRS, 75005 Paris, France}
\email{wilfrid.poirier@lne.fr}
\date{\today}

\begin{abstract}
Measurements of heat flux between micro-objects, in vacuum or in air, are challenging because of their small size and the low thermal conductance 
of the medium between them. One way to address this issue consists in using a scanning thermal microscope (SThM) equipped with a temperature dependent 
resistance thermometer. However, this requires an instrument able to both injecting a defined heating Joule power and performing highly-sensitive resistance 
measurements. Here, we present such an instrument based on a Wheatstone bridge equipped with three Kelvin arms. It can perform resistance measurements in the range 
from 100 $\Omega$ to 1000 $\Omega$ not only in direct current but also in alternating current regimes at frequencies up to a few tenths of kHz. We first show 
that measurements of resistance standards are accurate to within one part in $10^4$ with a relative experimental standard deviation which can be as low as 
one part in $10^8$ for one second measurement. The instrument is then tested with a SThM thermometer. With the support of an electro-thermal model considering thermal 
time constants of the thermometer, we explain the frequency dependence of detected signals and optimize the measurement protocols of temperature and heat flux. 
By measuring sub-mK temperature variations, this instrument is then used to determine with a few nanowatts uncertainty the near-field radiative heat flux between 
a heated glass microsphere and a glass substrate, which is caused by the coupling of surface phonon-polaritons. 
\end{abstract}
\maketitle

\section{Introduction}
Scanning thermal microscope (SThM) is an atomic force microscope (AFM) combining a high-precision scanning system and a cantilever equipped with a thermal sensor.
It enables the measurement of temperature and the assessment of heat flux at the nanoscale, for fundamental physics investigations and industrial applications of 
nanostructures or nanomaterials \cite{Zhang2020,Menges2016,Gomes2015}. The most widely used thermal sensors are based on the bending of a bimaterial cantilever \cite{Rousseau2009}, 
the Seebeck effect in a thermocouple \cite{Kittel2005} and the resistance variation of a temperature-dependent filament \cite{Menges2016,Bourgeois2007}. Among them, 
resistance-based thermal sensors are particularly attractive because they can operate both as thermometers and local heaters through controlled Joule dissipation. 
Regrettably, no SThM manufacturer currently provides instrumentation capable of measuring the probe thermometer resistance with the accuracy and sensitivity required 
to investigate either low-thermal-conductivity materials or radiative heat transfer between micrometer-sized objects in vacuum\cite{Song2015,Kittel2008,Yan2023}. 
On the other hand, various measurement techniques \cite{Delahaye1992,Trapon1997} have been developed by national metrology institutes (NMIs) to perform accurate and sensitive resistance measurements, 
which are particularly effective at detecting small variations in electrical resistance standards.These techniques can be leveraged to improve the accuracy and precision of resistance thermometry. 
One consists in comparing voltages at terminals of both the resistive thermometer and a highly-stable resistance standard connected in series using four-wire differential 
voltage measurements\cite{Delahaye1992,Trapon1997}. This method has the advantage to be insensitive to contact resistances, but requires an accurate measurement of the ratio (or difference) of two voltages. 
The usual technique consists in performing the comparison of the two successive voltage measurements by using an auxiliary potentiometer associated with a null detector, which 
is efficient but complicates the experiment. Ratio-bridges based on current comparators have represented an important advance in resistance metrology. The two resistors 
to compare are supplied with currents of values ensuring a voltage balance. The resistance ratio is therefore obtained from the current ratio, which is determined itself 
from the ratio of numbers of turns of windings of a transformer (windings are wound around a high magnetic permeability core). Commercial bridges of that type are the most 
accurate provided that currents are high enough. But, they are very expensive so that their use is restricted to measurement of electrical standards in NMIs. Ratio-bridges based on a cryogenic current comparator\cite{Poirier2020} allows comparison of resistances with a relative uncertainty of few parts in $10^{10}$. 
One simpler and more cost-effective alternative is the Wheatstone bridge. Thus, it is widely used for SThM applications in physics\cite{Lefevre2005} and 
metrology \cite{Fleurence2023, campbell2025}. It allows a direct opposition of voltages at terminals of the thermometer and of a resistance standard. Close to equilibrium, 
the accuracy of the biasing voltage applied to the bridge and of the unbalance voltage measurement are not crucial. Only a sensitive differential amplifier is required to 
measure the unbalance voltage. On the other hand, resistances in the Wheatstone bridge are usually defined with two wires, which is acceptable only for low resistance 
values of wires compared to that of resistors. However, approximate four-definition of resistances in a Wheatstone bridge can be achieved by integrating additional Kelvin arms, 
turning it into a Thomson\cite{Trapon1997} or Warshsawky bridge\cite{Warshawsky1955, Delahaye1992}. Such Wheatstone bridge types are also used in NMIs for performing accurate 
comparisons of resistance standards. The universality of the quantum Hall effect was even checked with the record relative uncertainty of 3 parts in $10^{11}$ using a specific 
Wheatstone bridge made of quantum resistance standards\cite{Schopfer2007, Schopfer2013}. Here, we report on a versatile Wheatstone bridge equipped with three Kelvin 
arms able to measure resistance in the range from 100 $\Omega$ to 1000 $\Omega$ both in direct current (DC) and alternating current (AC) regimes. 
Measurements of resistance standards are accurate to within one part in $10^4$ with a relative experimental standard deviation which can be as low 
as one part in $10^8$ for one second measurement. With the support of an electro-thermal model of a SThM equipped with a platinum-film thermometer, 
we optimize the measurement protocols to achieve the measurement of sub-mK temperature variations while injecting a defined Joule power in the thermometer. 
This allows us to measure near-field radiative heat flux with a few nanowatts uncertainty resulting from surface phonon-polariton coupling between a heated 
glass microsphere and a glass substrate.
\section{Principle of the measuring instrument}
\subsection{The resistance bridge}
\begin{figure}[h]
\centering
\includegraphics[width=3.4in]{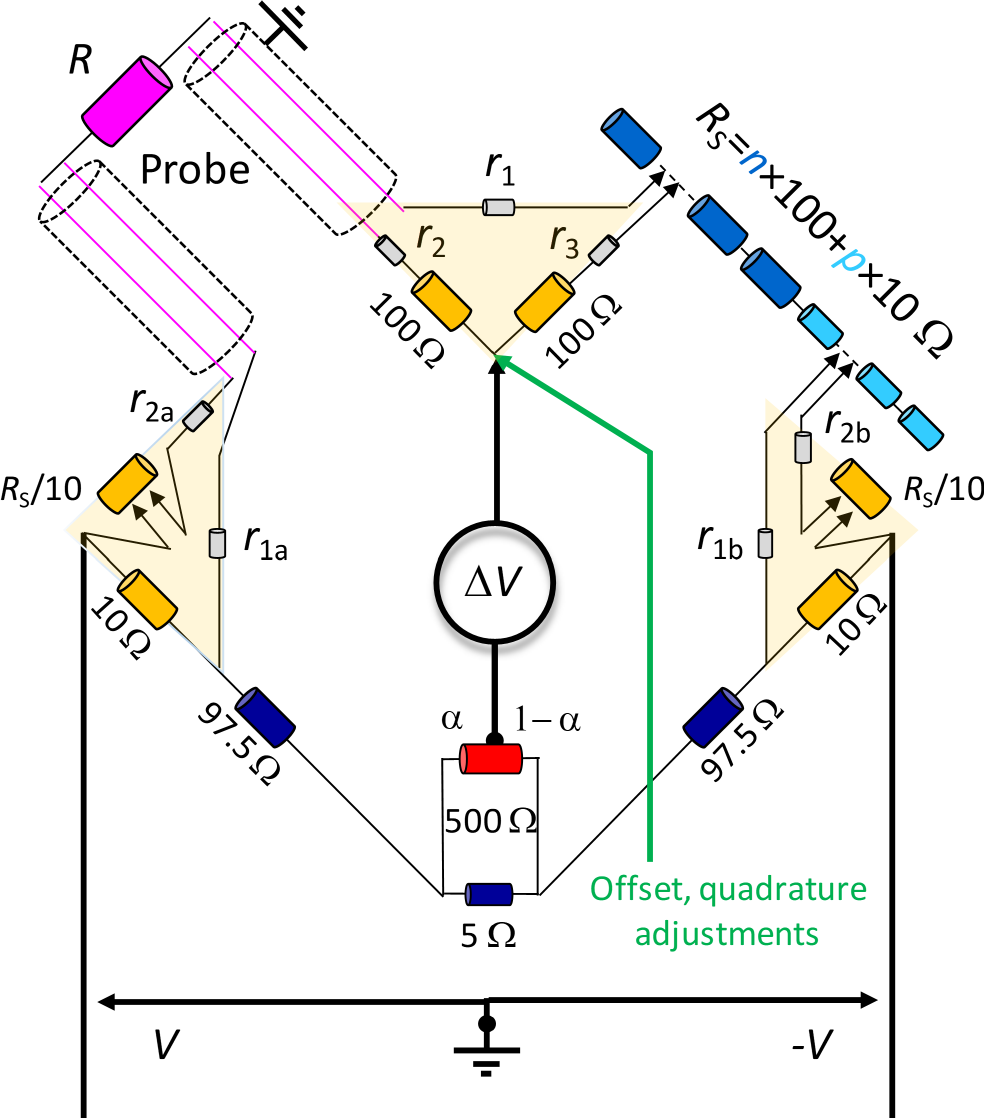}
\caption{Schematic of the resistance bridge, which includes the electrical resistance, $R$ (magenta), of the thermometer and the adjustable reference standard resistance, $R_\mathrm{S}$, at the top (sky blue for $100~\Omega$ resistors, turquoise blue for $10~\Omega$ resistors), and two equivalent fixed resistances of 100 $\Omega$ (97.5 + 5/2 $\Omega$) at the bottom (dark blue). The Wheatstone bridge is equipped with three Kelvin arms (orange-shaded triangle). A potentiometer of 500 $\Omega$ in parallel with a 5 $\Omega$ resistance allows fine voltage balance of the bridge. $r_1$, $r_2$, $r_3$, $r_{1a}$, $r_{2a}$, $r_{1b}$ and $r_{2b}$ represent parasitic resistances of the thermal probe wires and of contacts with $R_\mathrm{S}$. Offset and quadrature signal adjustments can be injected at the top terminal of the null detector (green). The bridge is symmetrically biased by $V$/-$V$ voltages. Dashed-line cylinders represent grounded shieldings of the wires connecting the thermometer of resistance $R$.}\label{fig1}
\end{figure}
Fig.\ref{fig1} shows a detailed schematic of the resistance bridge. It is a Wheatstone bridge equipped with three Kelvin arms, which are used to allow an approximate four-wire definition of the resistances. It is made of the electrical resistance of the thermal probe, $R$, an adjustable reference standard resistance, $R_\mathrm{S}$, and two equivalent fixed resistances of 100 $\Omega$ (equal to 97.5 + 5/2 $\Omega$). The bridge is coarsely balanced through the adjustment of the $R_\mathrm{S}=n\times100+p\times10~\Omega$ value by setting $n$ and $p$ integers. The fine in-phase adjustment of the bridge is achieved setting a 500 $\Omega$ potentiometer connected in parallel to a 5 $\Omega$ resistance. Residual offset and quadrature unbalance voltage can be cancelled by injecting compensation signals at the top terminal of the null detector. The three Kelvin arms make the bridge equilibrium very insensitive to parasitic resistances of the thermal probe wires and of contacts with $R_\mathrm{S}$, represented in fig.\ref{fig1} by $r_1$, $r_2$, $r_3$, $r_{1a}$, $r_{2a}$, $r_{1b}$ and $r_{2b}$. This is explained by fig.\ref{figEWheatstone} of appendix A, which shows the equivalent circuit of the bridge after application of triangle-star transformations. $r_1$ resistance is equally shared between $R$ and $R_\mathrm{S}$ resistances to within a relative correction of $\mathcal{O}(r/200)$. In the same way, $r_{1a}$ ($r_{1b}$) is shared between $R$ ($R_\mathrm{S}$) and $100~\Omega$ resistances according to the ratio $\frac{R_\mathrm{S}}{100}$ to within a relative correction of $\mathcal{O}(10r/R_\mathrm{S})$. It results that the relative deviation, $\Delta R/R_\mathrm{S}$, of $R$ from $R_\mathrm{S}$ is sensitive to parasitic resistances only at the second order, which improves in a major way the reproducibility and accuracy of the resistance bridge. Moreover, the bridge is symmetrically biased by $V$/-$V$ voltages with a grounded mid-point. Low and high potential wires of the thermal probe are separately screened by a shielding at ground potential. Close to bridge equilibrium, low potential wires are at a potential close to ground. This strongly reduces circulation of quadrature leakage currents caused by wire capacitance, increasing frequency bandwidth of the measuring instrument.\\     
In absence of parasitic resistances, the nominal value of the thermal probe resistance $R$ is given by: 
\begin{equation}
\small
R=k_rR_\mathrm{S}\frac{[1-a(1-2\alpha)]-\gamma[\frac{120+bR_\mathrm{S}}{100+R_\mathrm{S}}]}{[1+a(1-2\alpha)]+\gamma[\frac{100+bR_\mathrm{S}}{100+R_\mathrm{S}}]},\label{eq1}
\end{equation}
\normalsize
where $k_r$ is the bridge ratio nominally equal to unity if both bottom resistors of nominal resistance 100 $\Omega$ are equal, $a$ and $b$ are nominally equal to $a=\frac{5//500}{\Sigma_\mathrm{R}}=2.47586\times10^{-2}$ and $b=1+\frac{20}{\Sigma_\mathrm{R}}=1.10002476$ respectively, with $\Sigma_\mathrm{R}=2\times 97.5+5//500=199.950495$ $\Omega$ (sum of bottom resistances), $\alpha \in$ [0,1] is the potentiometer setting and $\gamma=\Delta V/V$ is the relative unbalance voltage. The current flowing through the probe resistance is given by:
\scriptsize
\begin{equation}
I_\mathrm{probe}=\frac{2\Sigma_\mathrm{R}(100+R_\mathrm{S})V}{R[20R_\mathrm{S}+\Sigma_\mathrm{R}(100+R_\mathrm{S}]+R_\mathrm{S}[20R_\mathrm{S}+\Sigma_\mathrm{R}(120+R_\mathrm{S})]},\label{eq2}
\end{equation}
\normalsize
Let us note that $R_\mathrm{S}$, $k_r$, $a$ and $b$ can be calibrated for each measuring instrument using reference standard resistors. The deviation of the ratio bridge from unity is given by the relative resistance discrepancy between the two bottom resistors of $100~\Omega$ nominal resistance. They were adjusted so that $k_r$ is close to unity to within a few parts in $10^5$.

The efficiency of Kelvin arms in reducing the effect of parasitic resistances is tested by increasing each of the four wires connecting the standard resistance, $R$, by a significant resistance 
of $2~\Omega$. This results in shifting the estimation of the standard resistance, $\frac{\Delta R}{R_\mathrm{S}}$, by 
$2.1\times10^{-3}$, $1.8\times10^{-4}$ and $4.3\times10^{-5}$ for nominal standard resistance of $100~\Omega$, $500~\Omega$ and $1000~\Omega$, respectively. These measured shifts, which are very close to those predicted by assuming initial resistance for $r_1$, $r_2$, $r_3$, $r_{1a}$, $r_{2a}$, $r_{1b}$ and $r_{2b}$ below 0.1 $\Omega$ (see appendix A), clearly demonstrate second order rejection of parasitic resistances. According to equation (\ref{eqDR}) in Appendix A, a typical resistance value below $\sim 0.1$ $\Omega$ for each wire of the thermometer (i.e for $r_{1a}$, $r_{2a}$, $r_1$, $r_2$) would lead to a relative error in the measurement of $R$, $|\Delta R/R_\mathrm{S}|$, less than $5\times10^{-6}$ and $10^{-7}$ for $R_\mathrm{S}=100$ $\Omega$ and $R_\mathrm{S}=1000$ $\Omega$ respectively. Any supplementary variation of a parasitic resistance by 10 m$\Omega$, caused either by lack of reproducibility of commutator contacts or 
ambient temperature change, would result in very small relative resistance measurement shift of less than $\sim 5\times10^{-7}$ and $\sim 1\times10^{-8}$ for $R_\mathrm{S}=100$ $\Omega$ and $R_\mathrm{S}=1000$ $\Omega$, respectively. Considering a typical relative dependence on temperature for $R$ of $2.5\times10^{-3}/\mathrm{K}$ (as in \cite{Guillemot2025} with a platinum thermometer), this would correspond to temperature measurement shifts of 0.2 mK and 0.004 mK, respectively. The stability and the reproducibility of the resistance bridge also rely on the use of high-precision and high-stability Vishay resistors (Z201 series) characterized by a low nominal temperature coefficient of only 0.05 ppm/°C, which are soldered on commutators based on low resistance (typically less than 10 m$\Omega$) gold-plated silver contacts (from IEC Monaco company). 

To test the wire shielding for reducing the circulation of quadrature currents caused by parasitic capacitances, the bridge is biased with a sinusoidal voltage signal at a frequency $f$ (pulsation $\omega=2\pi f$) of 11.113 kHz. The deviation of the relative unbalance voltage, $\gamma$, is then measured when a large 33 nF capacitor is successively connected between one terminal of the resistance standard and the ground. Despite the low impedance of the capacitor at this frequency (about 434 $\Omega$), deviations of the imaginary part, $\gamma_y^\omega$, of $\gamma^\omega$, lower than $5\times10^{-4}$ are measured for a 1000 $\Omega$ resistance standard, demonstrating the strong rejection of quadrature leakage currents. We therefore expect small deviations of $\gamma_y^\omega$, of only $2\times10^{-6}$ for a typical cable capacitance of 100 pF at a frequency of 11 kHz. Given that the lock-in phasing angle is correctly adjusted to within 20 mrad, we do not expect deviations of the real part $\gamma_x^\omega$, of $\gamma^\omega$, larger than a few parts in $10^{-8}$. 

The sensitivity of the bridge is determined by the Johnson-Nyquist noise of the equivalent resistance, (225 $\Omega$ + $R_\mathrm{S}/2$), and by the noise of the unbalance voltage detector. \\
\subsection{Electronics and instruments}
\begin{figure}[h]
\centering
\includegraphics[width=3.4in]{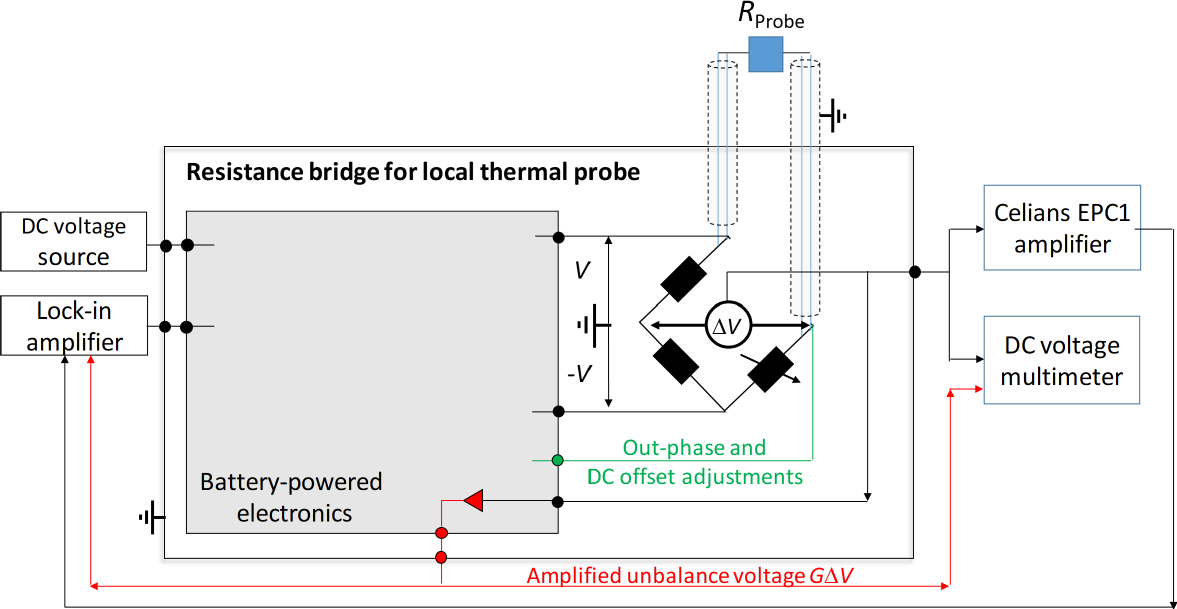}
\caption{Schematic of the components of the measuring instrument: the resistance bridge, the battery-powered electronics, the external DC and AC voltage sources, the DC and AC voltage meters for detecting the unbalance voltage of the bridge.}\label{fig2}
\end{figure}
Fig.\ref{fig2} shows the schematic of the measuring instrument. The resistance of the thermal probe is one element of a Wheatstone bridge, which is biased by symmetric voltage sources, $V$, $-V$ provided by a battery-powered electronic circuit. External DC and AC voltage sources can be used to control the biasing voltage using selectable gains. The electronic circuit also provide offset and quadrature signals, which can be injected at the top mid-point of the Wheatstone bridge to achieve perfect voltage balance. It also includes an internal low-noise and high-impedance differential amplifier (AInt), which can be used to amplify the unbalance voltage of the bridge to replace an external amplifier. The measuring instrument and the wires of the resistance thermal probe are carefully shielded at ground potential.
\begin{figure}[h!]
\centering
\includegraphics[width=3.4in]{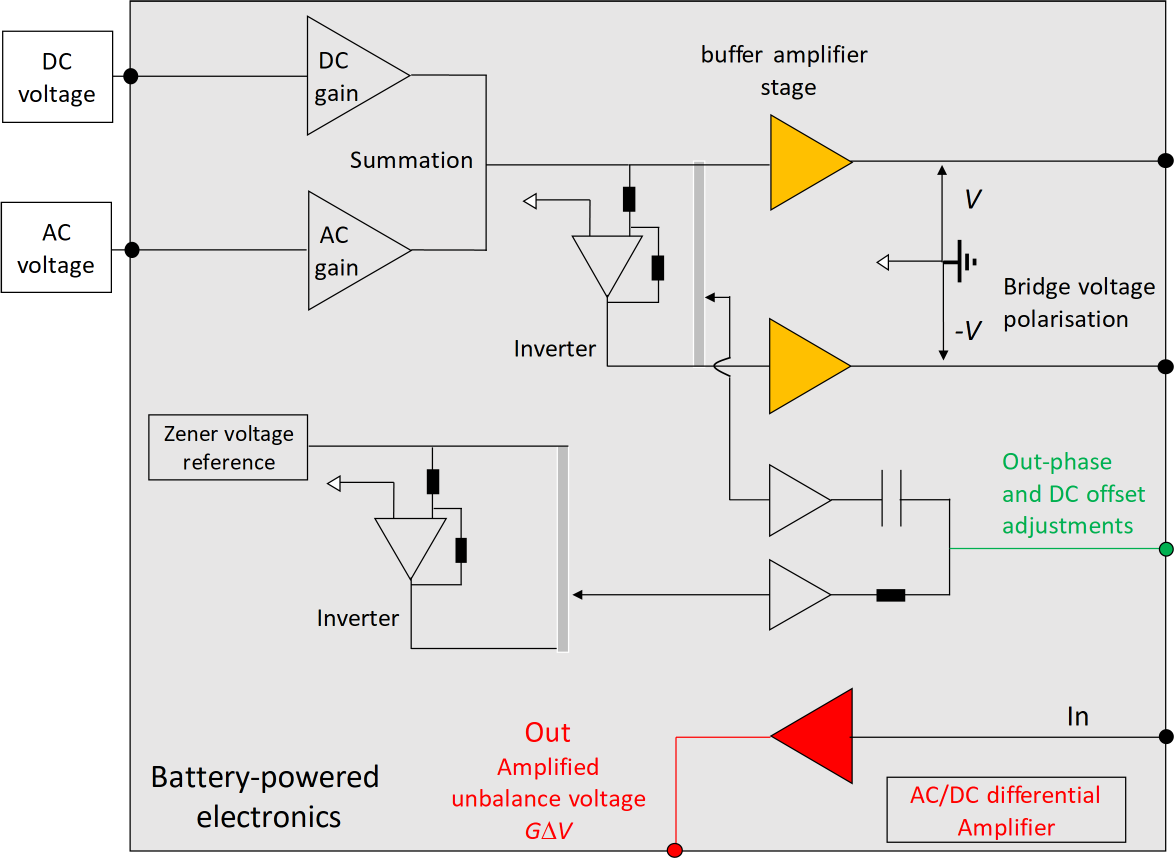}
\caption{Schematic of the bridge electronics. DC and AC external voltages are combined to generate symmetric voltage biasing the Wheatstone bridge. Offset and quadrature signals are prepared. 
The electronic card also includes a differential amplifier which can be used to amplify the unbalance voltage of the bridge.}\label{fig3}
\end{figure}

Fig.\ref{fig3} shows a simplified schematic of the electronic circuit. After electrical insulation and amplification using high-impedance instrumentation amplifiers (based on AD795 operational amplifier), external DC and AC control voltage signals are summed. An inverter is used to provide an opposite voltage. Two buffer amplifier stages are then used to produce a symmetric voltage biasing, $V$, $-V$, 
which is applied to the resistance bridge. Quadrature and DC offset signals are prepared to be injected in the bridge in order to improve the voltage balance. The resulting unbalance voltage can be measured using a DC multimeter or fed into a lock-in amplifier after amplification by a Celians EPC1 amplifier. It can also be amplified by the internal low-noise and high-impedance differential amplifier (AInt) before measurement. The unknow resistance value is then determined using Equation (\ref{eq1}). Electronics circuits are based on precision resistances from Vishay (S and Z series) characterized by very low nominal temperature coefficients (0.5 ppm/°C and 0.05 ppm/°C respectively) and precision operational amplifiers (ADA4625-1, OPA97, BUF634T). The main circuit board is powered by a supply card providing 15V/+15V voltages stabilized by a Zener voltage reference and boosted by power amplifiers (OPA547T). To reduce noise from environment, it is itself supplied by rechargeable lead batteries.
\begin{figure}[h!]
\centering
\includegraphics[width=3.4in]{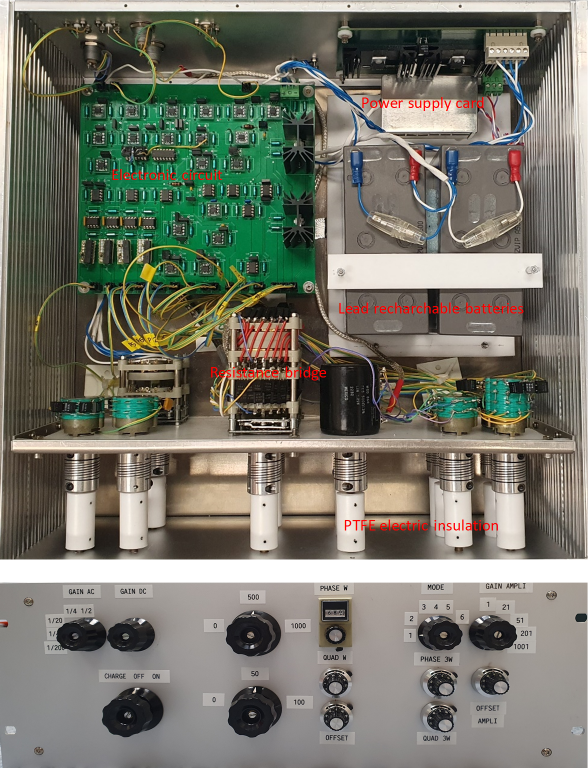}
\caption{Pictures of top and front views of the measuring instrument.}\label{fig4}
\end{figure} 

Fig.\ref{fig4} show pictures of top and front view of the measuring instrument. Electronic cards, batteries and commutators are packed in a metallic box while being electrically insulated from it. Resistances of the Wheatstone bridge, except that of the thermal probe, are soldered on two main IEC Monaco rotary commutators, which are used to vary the resistance value $R_\mathrm{S}$ by changing $n$ and $p$ integers (from 0 to 10). Others commutators are used to vary DC and AC voltage gains, measurement modes, amplifier gain. All commutators and potentiometers are electrically insulated by PTFE pieces to avoid hand-effects during operation.  Let us remark that the measuring instrument also includes compensation circuits to balance the $\omega$ and $3\omega$ signals caused by the AC self-heating of the thermal probe. This feature, designed to perform accurate and sensitive $3\omega$ measurements, is not described in this work. 
  
\section{Characterization of the instrument in measuring a resistance standard}
We first evaluated the performances of the resistance bridge for measuring resistance standards with negligible temperature dependence and values ranging from 100 $\Omega$ and 1000 $\Omega$.
\subsection{DC measurements}
From the relationship (\ref{eq1}), one can establish the Type B standard uncertainty budget (uncertainty evaluated using methods other than statistical ones) \cite{BIPMGUM} for resistance measurement in the DC regime. Main contributions are related to calibrations of resistance standard $R_\mathrm{S}$, resistance ratio $k_r$ of the bridge, $a$ and $b$ parameters, $\alpha$ adjustability and calibration of voltage ratio $\gamma$. Table \ref{Type-B budget} reports for each contributing quantity: its relative measurement uncertainty ($u$), its impact on the resistance determination (sensitivity) and its relative type B contribution ($u_\mathrm{B}$) to the resistance measurement. The contribution of the wires connecting the resistance standard is not considered in this budget. As discussed previously, one should use equation (\ref{eqDR}) to estimate the relative error (and its uncertainty) originating from the resistance of connection wires. Lowest uncertainties are achieved at equilibrium ($\gamma\simeq0$) and resistance value close to $R_\mathrm{S}$ ($\alpha\simeq 1/2$).  
\begin{table}[ht]
\caption{\textbf{Type B standard uncertainty budget of resistance measurement in the DC regime}\\ The coverage factor is $k=1$.}\label{Type-B budget}
\begin{center}
\begin{tabular}{|c|c|c|c|c|}
  \hline
 \textbf{Contributions}&$u$&Sensitivity&$u_\mathrm{B}$ ($\gamma=0$)\\ 
  &($10^{-6}$)& &($10^{-6}$)\\ \hline
  $R_\mathrm{S}$ calibration&10&1&10\\
  $kr$ calibration&5&1&5\\
  $a$ calibration&20&$2|1-2\alpha|$&$40|1-2\alpha|$\\
  $b$ calibration&200&$2\gamma\frac{R_\mathrm{S}}{100+R_\mathrm{S}}$&0\\
  $\alpha$ adjustability&120&$4a\simeq 10^{-1}$&12\\
  $\gamma$ calibration&negligible&2&$\sim 0$\\ \hline
  \textbf{Total}  & & &$16~(\alpha=1/2)<u_\mathrm{B}$\\ 
     & & &$<46~(\alpha=0;1)$\\ \hline
\end{tabular}
\end{center}
\end{table} 
\normalsize

\begin{figure}[h!]
\centering
\includegraphics[width=3.4in]{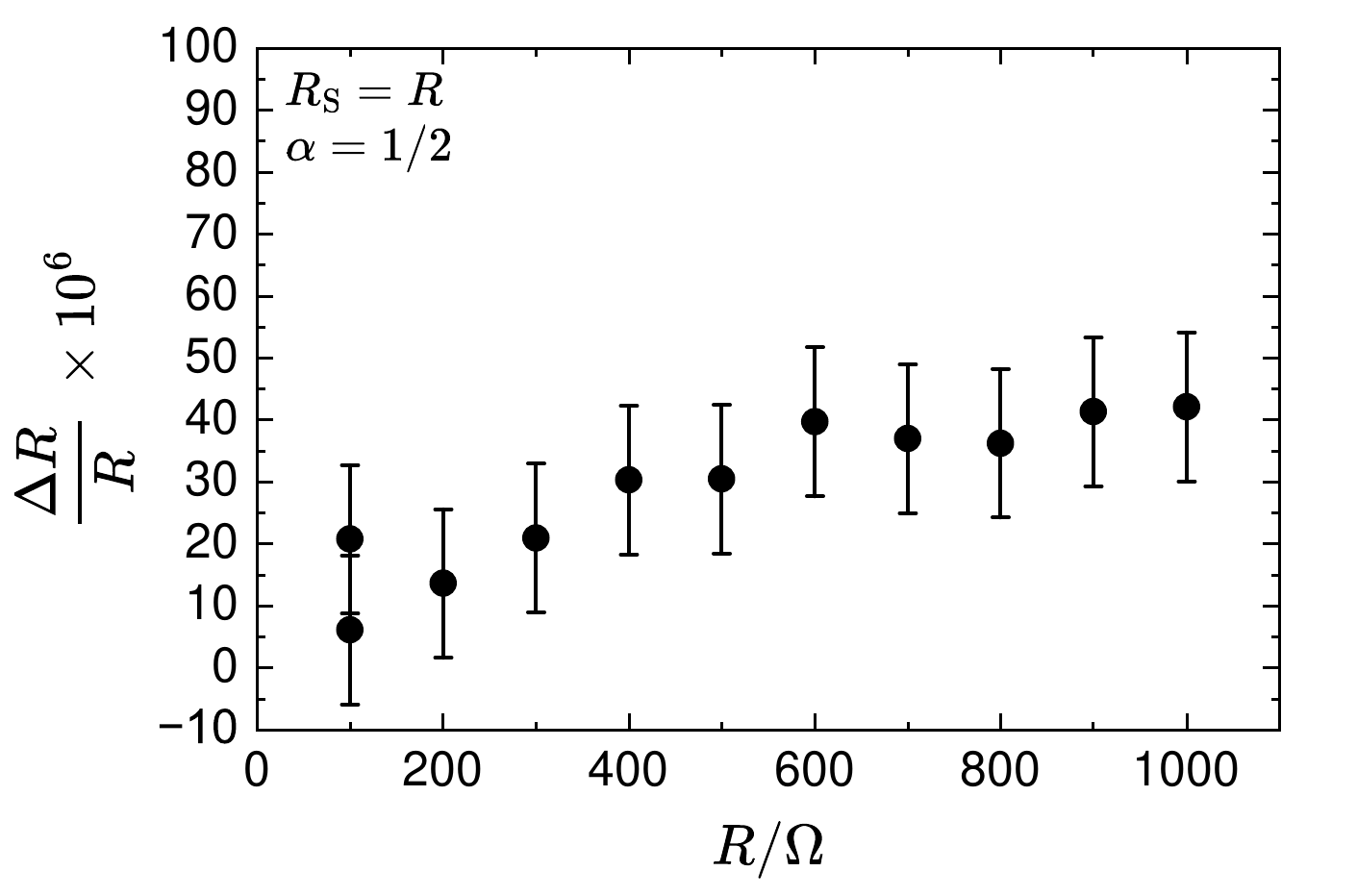}
\caption{Relative deviations of resistances (SR1010) from their calibrated values. Uncertainties are obtained by summing all contributions, except those of $R_\mathrm{S}$ and resistance ratio $k_r$.}\label{ExactitudeDC}
\end{figure}

To estimate Type B contributions related to the  calibration of resistance standard $R_\mathrm{S}$ and of the resistance ratio $k_r$ of the bridge, calibrated ESI SR1010 resistors in the range from 100 $\Omega$ to 1000 $\Omega$ were measured using the measuring instrument. Fig.\ref{ExactitudeDC} reports the relative deviation between the measured value (assuming that $k_r=1$ and $R_\mathrm{S}$ is at nominal value) and the calibrated value. It shows positive discrepancies ranging from a few $10^{-6}$ up to about $40\times10^{-6}$, with a mid deviation of $25\times10^{-6}$. Without applying any correction and using nominal value of parameters, the bridge is accurate to within 50 parts in $10^{5}$. Observed deviations can be explained by the combination of an imperfect ratio, $k_r$, of the bridge deviating from unity by $\sim-25\times10^{-6}$ (this causes a deviation independent of the resistance value) and of relative discrepancies of $R_\mathrm{S}$ values from their nominal values by less than $\sim 20\times10^{-6}$. Consideration of $R_\mathrm{S}$ and $k_r$ calibrations leads to type B relative uncertainties ranging from $16\times 10^{-6}$ ($\alpha=1/2$) to $46\times 10^{-6}$ ($\alpha=0;1)$), lower than 50 parts in $10^6$ (see table \ref{Type-B budget}).  
\subsection{AC measurements}
\begin{figure}
\centering
\includegraphics[width=3.4in]{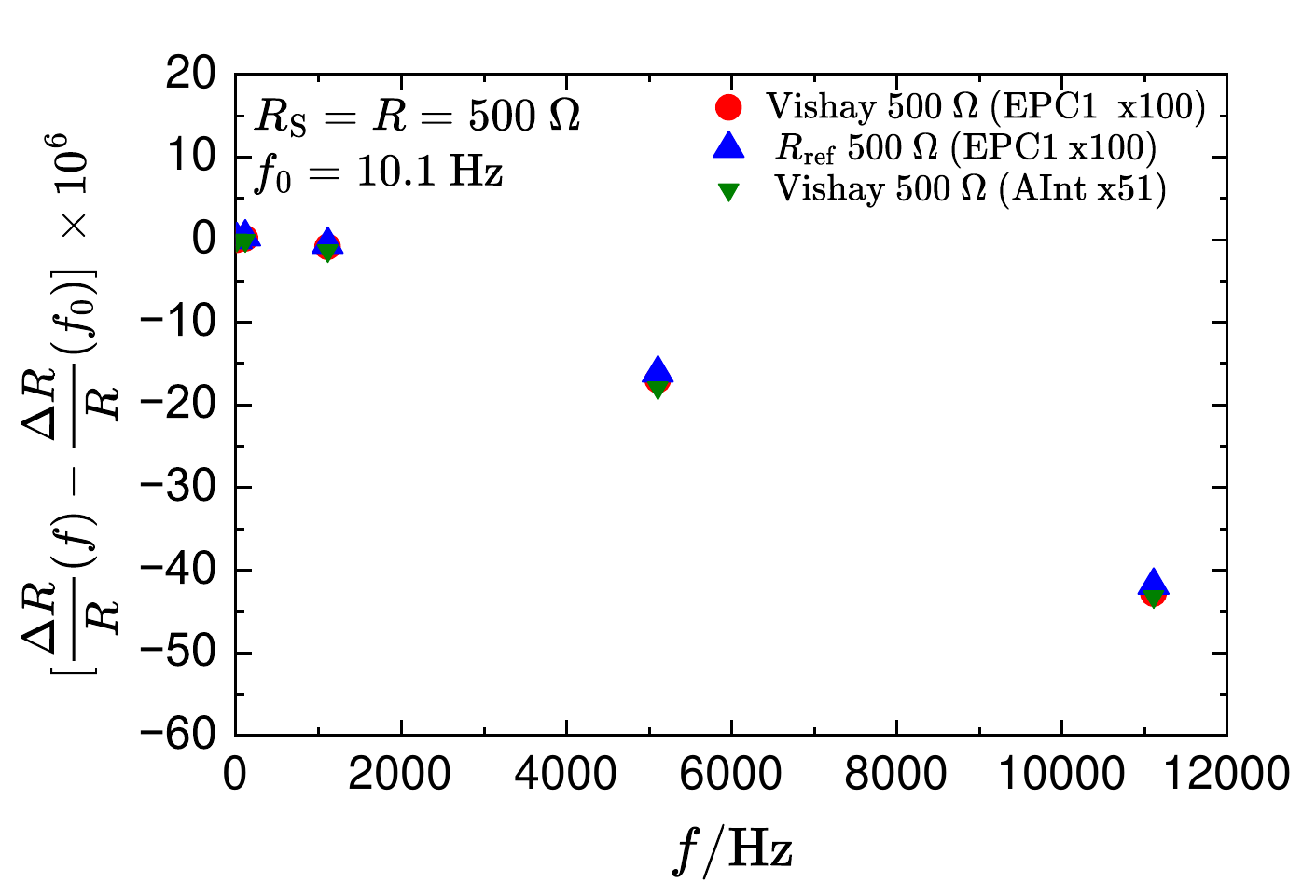}
\caption{Difference between the relative resistance deviations measured at frequencies $f$ and $f_0=10.1$ Hz. Measurements are carried out with a 500 $\Omega$ Vishay resistor using either the Celians EPC1 amplifier (filled red circle) or the internal amplifier (green triangle), and a calibrated low-frequency dependence 500 $\Omega$ resistor using the EPC1 amplifier (blue triangle).}\label{ExactitudeAC}
\end{figure}

Fig.\ref{ExactitudeAC} shows the frequency dependence of resistance measurements performed with two types of reference resistors and using either the Celians EPC1 amplifier or the internal amplifier AInt. The frequency dependence is the same independently of resistors and amplifier used showing that it is specific to the bridge itself. A deviation of less than 50 parts in $10^6$ is measured at about 11 kHz.
One can therefore expect relative errors below 100 parts in $10^6$ in the measurement of a resistance in the range from 100 $\Omega$ to 1000 $\Omega$ at frequencies up to 10 kHz. This accuracy level in the resistance measurement is beyond the requirement of the thermal nanometrology field because local temperature and thermal conductivity measurements generally involve many other physical parameters leading to a total measurement uncertainty larger by orders of magnitude \cite{Fleurence2023} (thermal probe calibration, contact thermal resistance, existence of water meniscus in air...).

\subsection{Noise level an stability}
Others crucial characteristics are the noise level and stability of the measuring bridge. At equilibrium ($\gamma=0$), the voltage noise detected after amplification is expected to only result from the combination of the Johnson-Nyquist noise of the resistors constituting the bridge and of the amplifier voltage noise.    
\begin{figure}[h!]
\centering
\includegraphics[width=3.4in]{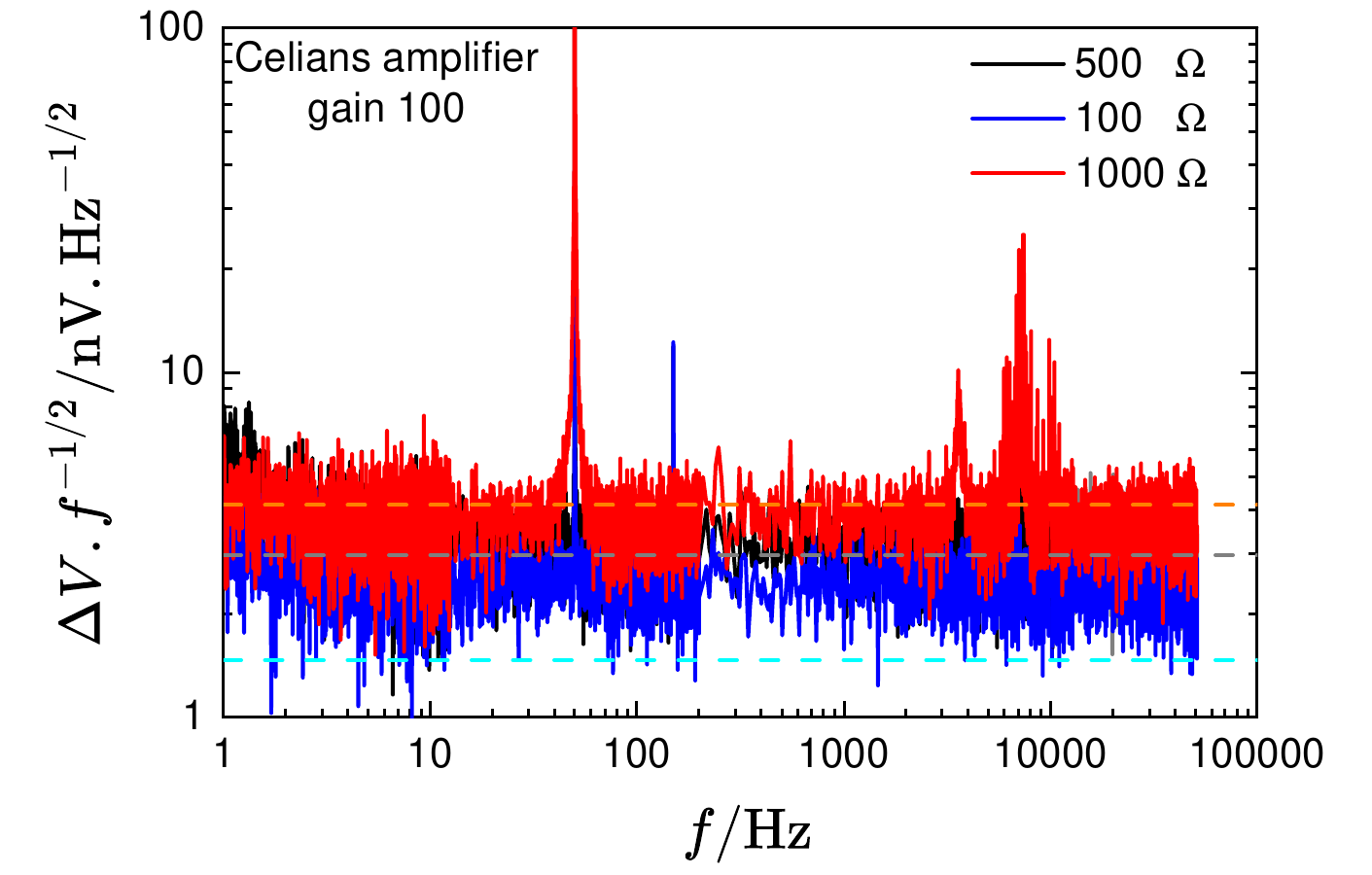}
\caption{Voltage noise spectral density detected at the terminals of the bridge using a Celians EPC1 amplifier with 100 gain for the measurement of 100 $\Omega$ (blue), 500 $\Omega$ (black) and 1000 $\Omega$ (red) resistance. Dashed lines represent calculated noise levels.}\label{Spectre Celians}
\end{figure}
Fig. \ref{Spectre Celians} shows the noise spectral density of $\Delta V$ measured using a Celians EPC1 amplifier with gain 100 as a function of frequency $f$ for the Wheatstone bridge balanced with 100 $\Omega$, 500 $\Omega$ and 1000 $\Omega$. From 1 Hz up to 100 000 Hz, except for a few peaks (notably 50 Hz) spectra are flat indicating that white noise dominates. Base noise levels are about of 2 $\mathrm{nV/Hz^{1/2}}$, 3 $\mathrm{nV/Hz^{1/2}}$ and 4 $\mathrm{nV/Hz^{1/2}}$, respectively. They agree well with the expected levels calculated by considering the Johnson-Nyquist noise of the bridge $(225+R_\mathrm{S}/2)$ and the amplifier noise (0.7 $\mathrm{nV/Hz^{1/2}}$). 
Type A uncertainty (uncertainty evaluated using statistical methods) \cite{BIPMGUM} of $\Delta R/R_\mathrm{S}$ is given by:
\begin{equation}
u_\mathrm{A}(\Delta R/R_\mathrm{S}) = 2 u_\mathrm{A}(\gamma) 
\end{equation}
For a biasing voltage $V=1$ V and an acquisition time of one second, one can therefore expect $u_\mathrm{A}(\Delta R/R_\mathrm{S})$ contributions from the bridge of 4, 6 and 8 parts in $10^9$ for $R_\mathrm{S}$ resistance of 100 $\Omega$, 500 $\Omega$ and 1000 $\Omega$, respectively. Fig. \ref{BruitMeasure} reports on the relative variation of a 500 $\Omega$ nominal resistance as a function of time, measured using external and internal amplifiers with a 1 V biasing AC voltage at a frequency of about 11 kHz and an acquisition time of 200 ms. The experimental standard deviation (per sample) amounts to  $9.3\times10^{-9}$ and $1.1\times10^{-8}$ using EPC1 and internal amplifier AInt, respectively, which is in agreement with the expectations. We can then extrapolate this result determined with fixed reference resistors to temperature measurements using a resistance thermometer. Considering the platinum thermometers on SThM probe used in this work, which are characterized by a typical resistance of $300~\Omega$ and a sensitivity of $(1.48\pm 0.05)~\mathrm{K}/\Omega$, we can therefore expect a contribution from the bridge itself to the type A temperature uncertainty as low as $2.2~\mathrm{\mu K}$ for one second measurement.  
\begin{figure}[h!]
\centering
\includegraphics[width=3.4in]{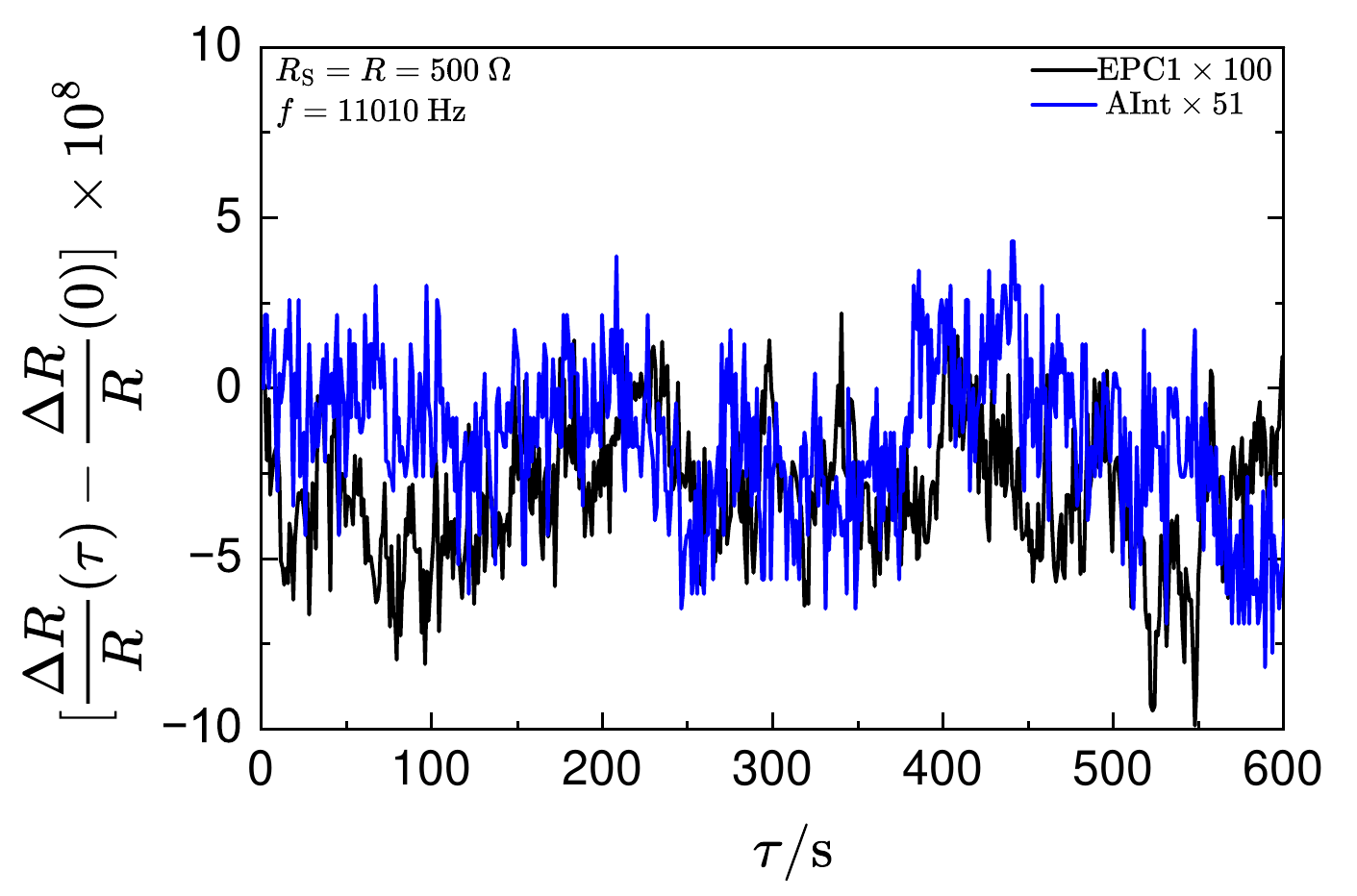}
\caption{Relative variation of the resistance, of $500~\Omega$ nominal value, as a function of time $\tau$, measured both with external Celians EPC1 amplifier (black) and internal amplifier (blue).} \label{BruitMeasure}
\end{figure}

\section{Measurement of local temperature and heat flux using a SThM resistance thermometer}
The determination of a local heat flux using a SThM thermometer requires injecting a known Joule power in the resistance, $P_\mathrm{J}=RI^2$, and measuring the temperature increase, $\Delta T$. The latter can be obtained from the measurement of the resistance variation, $\Delta R$, of the thermometer, which is strongly dependent on temperature. 
\begin{figure}[h!]
\centering
\includegraphics[width=3.4in]{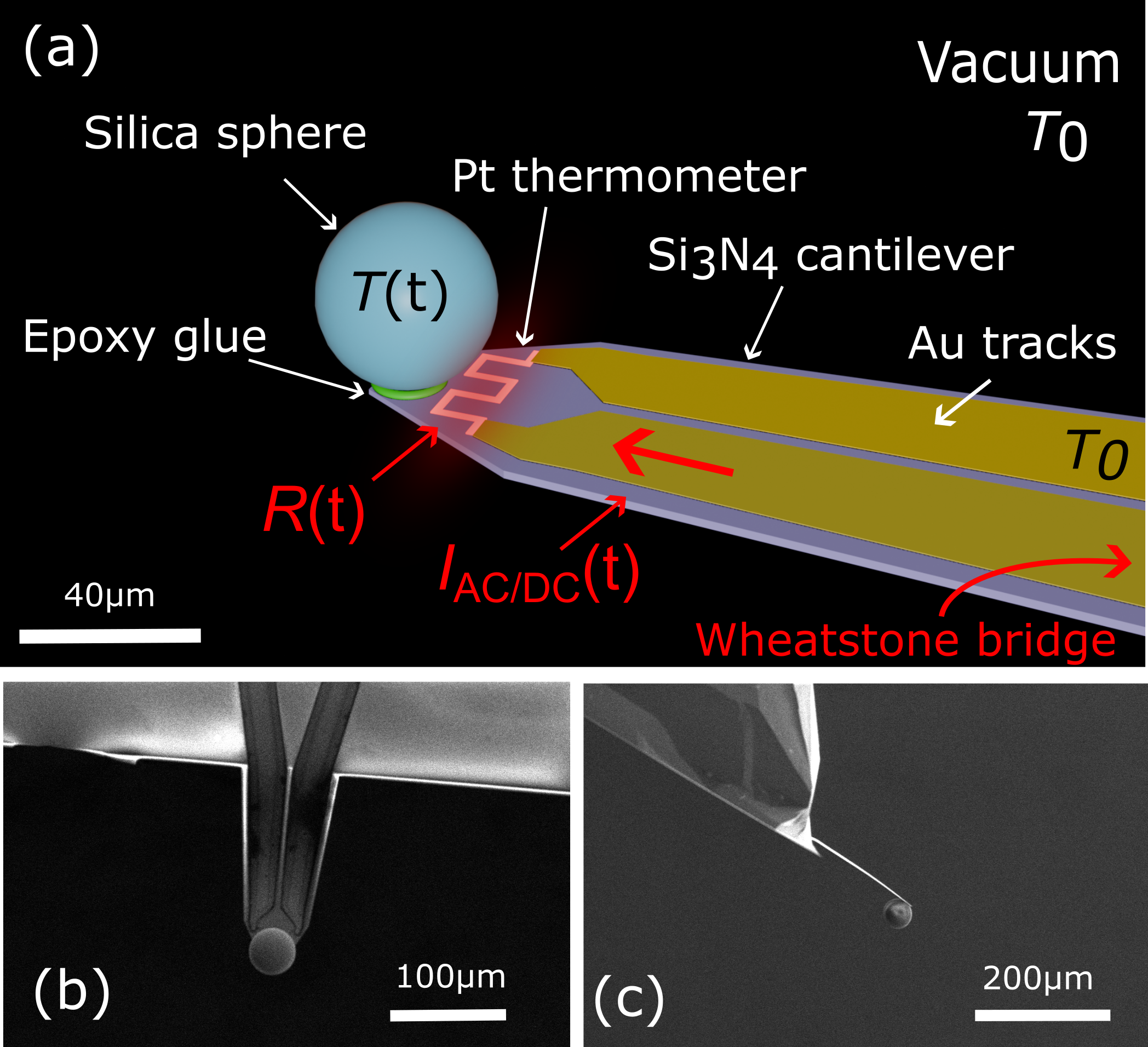}
\caption{Representation of the modified SThM probe used in the experiments. a) Schematic of the probe where the constitutive parts are labeled in white. b) and c) Pictures of the microsphere glued at the end of the cantilever close to the platinum serpentine.}\label{fig:figvic1} 
\end{figure}
Fig.~\ref{fig:figvic1}a) shows a schematic of the SThM thermometer, fabricated by Kelvin Nanotechnology Ltd (KNT), consisting of a $180~\mathrm{\mu m}$ long flat Si$_3$N$_4$ cantilever at the end of which is deposited a platinum serpentine of electrical resistance $R$. These thermometers have a typical resistance value, of 300 $\Omega$ at 300 K, which varies linearly on the temperature according to a 0.7 $\Omega$/K coefficient. Details can be found in \cite{Guillemot2025}. The base of the cantilever is maintained at 
ambient temperature $T_0$, while the thermometer is at temperature $T$. Fig.~\ref{fig:figvic1}a) also shows a borosillicate sphere of 42.3 $\mu$m diameter glued at the end of the cantilever close to the serpentine, which can be used to measure the radiative heat transfer in the sphere-plan geometry in vacuum (see next section). Figures \ref{fig:figvic1} b) and \ref{fig:figvic1}c) represent two scanning electron microscopy images of the probe from two different points of views. The four-wire connection of the thermometer to the Wheatstone bridge extends to the base of the cantilever. Its effectiveness in reducing the impact of the parasitic resistances is limited to those of the long wires connecting the cantilever to the measurement bridge. Beyond, the two-wire resistance is defined by the series resistance of the serpentine and of the gold tracks (typically $20~\Omega$ over the $300~\Omega$).

In this work, we have implemented the two following measurement protocols to determine the  temperature increase as a result of the injection of a known Joule power:  i) Biasing the Wheatstone bridge with a large DC voltage, $V_\mathrm{DC}$, and measuring the DC unbalance voltage, $V_\mathrm{AB}^\mathrm{DC}$. Equations (\ref{eq1}) and (\ref{eq2}) can be used to determine the resistance, $R$, and the current, $I$, which directly give the power value and the temperature $T$. However, the sensitivity can be limited by voltage offset drifts and large noise level of voltage detectors at low frequency;
ii) Biasing the Wheatstone bridge with a large AC voltage, $V_\mathrm{AC}$ and measuring the AC unbalance voltage, $V_\mathrm{AB}^\mathrm{AC}$. The mean Joule power due to the AC signal rises the temperature of the resistance and therefore its value to $R_\mathrm{AC}$. At frequency $f$ lower than the thermal frequency cutoff, the resistance, $R$, has an oscillating component at frequency $2f$ with a certain amplitude, $\beta$. This oscillating term introduces additional components to the unbalance voltage at frequencies $f$ and $3f$, which superimpose to the main signal making the relationship (\ref{eq1}) no longer valid. On the other hand, at frequency higher than the thermal frequency cutoff of the thermometer, the resistance and the therefore the temperature no longer oscillate. It results that their values can be determined using the relationship (\ref{eq1}), with an excellent statistical uncertainty because of the large AC signal used. 

\subsection{Electro-thermal model of the SThM resistance thermometer biased by an AC voltage polarization}
\begin{figure}[h!]
\centering
\includegraphics[width=3.4in]{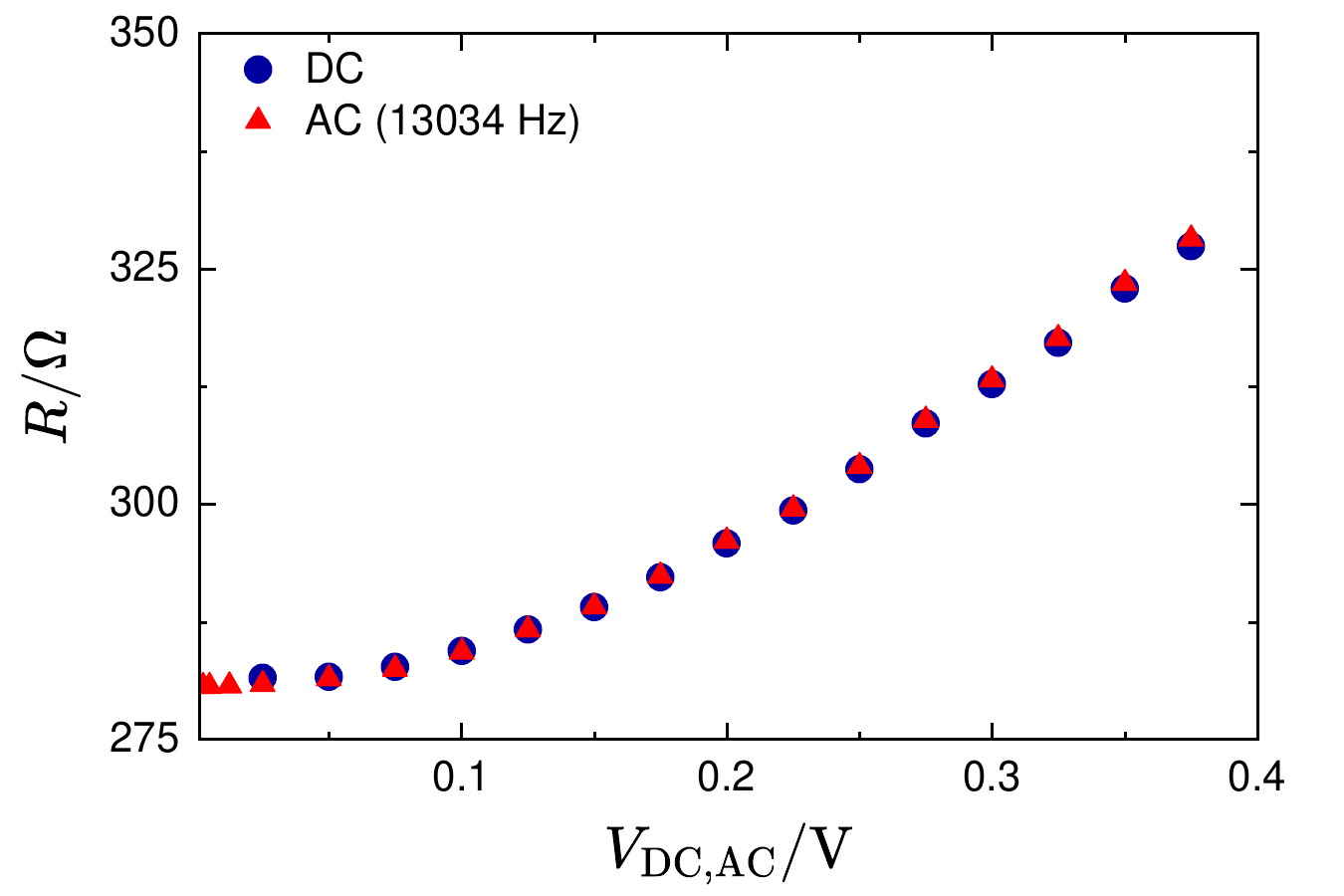}
\caption{Resistance $R$ determined from equation (\ref{eq1}) from measurement performed using either a DC bridge voltage $V_\mathrm{DC}$ or an AC bridge voltage $V_\mathrm{AC}$ at 13 kHz frequency.}\label{RDCAC}
\end{figure}

The SThM resistance thermometer in its thermal environment and connected to the measurement system can be described by a simple model considering a temperature $T$, an injected Joule power $P_\mathrm{J}$ and a complex conductance depending on pulsation $G(\omega)$ between the temperature $T$ and the environment temperature $T_0$. At the first order approximation, the current, $I_\mathrm{AC}(t)=I_\mathrm{AC}(\omega)\cos(\omega t)$, flowing through the resistance is proportional to the voltage applied the Wheatstone bridge $V_\mathrm{AC}(t)=V_\mathrm{AC}(\omega)\cos(\omega t)$ according to the relationship (\ref{eq2}). It results that the Joule power is given by:
\begin{equation}
P_\mathrm{J}=RI^2=\frac{RI_\mathrm{AC}(\omega)^2}{2}[1+\cos(2\omega t)],
\end{equation}
where $R$ is given by $R_0+\lambda\Delta T$, $R_0$ is the resistance at temperature $T_0$ (without Joule power) and $\lambda$ is the temperature coefficient in $\Omega$/K of the thermometer.
The mean Joule power, $\frac{RI_\mathrm{AC}(\omega)^2}{2}$, gives rise to a temperature elevation $\Delta T_0$. It is approximatively given by:
\begin{equation}
\Delta T_0=\frac{R_0I_\mathrm{AC}(\omega)^2}{2G(0)}\frac{1}{1-\frac{\lambda I_\mathrm{AC}(\omega)^2}{2G(0)}},
\end{equation}
It results a resistance increase from $R_0$ up to $R_\mathrm{AC}$ given by:
\begin{equation}
R_\mathrm{AC}=\frac{R_0}{1-\frac{\lambda I_\mathrm{AC}(\omega)^2}{2G(0)}},
\end{equation}
The second component of the Joule power at pulsation $2\omega$ leads to an oscillatory component of the temperature given by:
\begin{equation}
\Delta T'=\frac{R_\mathrm{AC}}{2|G(2\omega)|}I_\mathrm{AC}(\omega)^2\cos(2\omega t-\varphi),
\end{equation}
where the dephasing angle, $\varphi$, is given by:
\begin{equation}
\varphi=\arctan{\frac{Im(G(2\omega))}{Re(G(2\omega))}},
\end{equation}
From the total temperature elevation $\Delta T=\Delta T_0+\Delta T'$, one can calculate the expression of the resistance $R$ of the thermometer, which is given by: 
\begin{equation}
R=R_\mathrm{AC}[1+|\beta|\cos(2\omega t-\varphi)],\label{eqR}
\end{equation}
where $\beta$ is given by:
\begin{equation}
\beta=|\beta|e^{-i\varphi}=\frac{\Delta R_\mathrm{AC}}{R_\mathrm{AC}}\frac{G(0)}{G(2\omega)},\label{eqbeta}
\end{equation}
where $\Delta R_\mathrm{AC}=R_\mathrm{AC}-R_0$ is the resistance increase due to the mean Joule power. This relationship is the first order development in $\beta$, which is a valid approximation either for  $\Delta R_\mathrm{AC}/R_\mathrm{AC}\ll 1$ (moderate AC voltage) or for large pulsation $\omega$ above the thermal cutoff pulsation so that $\frac{|G(0)|}{|G(2\omega)|}\ll 1$. Let us note that $\lim\limits_{\omega \rightarrow +\infty} \frac{|G(0)|}{|G(2\omega)|}\rightarrow 0$. At very large pulsation, the resistance value and the temperature of the thermometer no more oscillate, and therefore are constant. Relationships (\ref{eq1}) and (\ref{eq2}) determined in DC regime are therefore valid at high frequency above the thermal cutoff frequency. This is confirmed by fig.\ref{RDCAC}, which shows that the resistance values deduced from measurements performed in DC and in AC at 13 kHz frequency are in agreement.
On the other hand, the amplitude of the resistance oscillation is maximal at $\omega=0$, for which $|\beta|=\frac{\Delta R_\mathrm{AC}}{R_\mathrm{AC}}$. In case of a simple thermal model for the conductance described by $G(\omega)=G_0+i\omega C_0$, $|\beta|$ is given by:
\begin{equation}
|\beta|=\frac{\Delta R_\mathrm{AC}}{R_\mathrm{AC}}\frac{1}{\sqrt{1+4\omega^2(C_0/G_0)^2}}
\end{equation}
This simple model takes into account one single thermal cutoff frequency given by $f_0=G_0/(4\pi C_0)$. 
\subsection{Relationship between unbalance voltage signals and thermometer resistance in AC regime}
The general expression of the relative unbalance voltage, $\gamma$, of the Wheatstone bridge is given by:
\begin{equation}
\gamma=\frac{R_\mathrm{S}c-Rc'}{DR_\mathrm{S}+D'R},\label{eqgamma}
\end{equation}
where $c=1-a(1-2\alpha)$, $c'=1+a(1-2\alpha)$, $D=\frac{120+bR_\mathrm{S}}{100+R_\mathrm{S}}$ and $D'=\frac{100+bR_\mathrm{S}}{100+R_\mathrm{S}}$ (here, we assume $k_r$=1). 
The continuous and $2\omega$ components of $R$ results in $\omega$ and $3\omega$ components for $\gamma$. 
From the expression (\ref{eqR}) of the resistance $R$, one can obtain the in-phase and out-phase components of $\gamma$ at 
frequency $\omega$ and $3\omega$ using the first approximation in $\beta$ (see Appendix B). Close to in-phase equilibrium at pulsation $\omega$ (i.e. $\gamma^{\omega}_x\simeq 0$), one finds:
\tiny
\begin{eqnarray}
\gamma^{\omega}_x &=&\frac{1}{DR_\mathrm{S}+D'R_\mathrm{AC}}[(R_\mathrm{S}c-R_\mathrm{AC}c')-\frac{Re(\beta) R_\mathrm{AC}c'}{2}]\\
\gamma^{\omega}_y &=&\frac{1}{DR_\mathrm{S}+D'R_\mathrm{AC}}[\frac{Im(\beta) R_\mathrm{AC}c'}{2}]\\
\gamma^{3\omega}_x &=&\frac{1}{DR_\mathrm{S}+D'R_\mathrm{AC}}[-\frac{Re(\beta) R_\mathrm{AC}c'}{2}]\\
\gamma^{3\omega}_y &=&\frac{1}{DR_\mathrm{S}+D'R_\mathrm{AC}}[\frac{Im(\beta) R_\mathrm{AC}c'}{2}]
\end{eqnarray} 
\normalsize
In-phase balance of the bridge at the pulsation $\omega$, i.e. $\gamma^{\omega}_x=0$, is experimentally obtained by adjusting $\alpha$. This corresponds to $(R_\mathrm{S}c-R_\mathrm{AC}c')-\frac{Re(\beta) R_\mathrm{AC}c'}{2}=0$. One obtains a dependence of $\alpha$ on pulsation $\omega$ given by:
\begin{equation}
\alpha(\omega)=\frac{1}{2}+\frac{1}{2a}\frac{R_\mathrm{S}-R_\mathrm{AC}-\frac{R_\mathrm{AC}}{2}Re(\beta)}{R_\mathrm{S}+R_\mathrm{AC}+\frac{R_\mathrm{AC}}{2}Re(\beta)} \label{eqalpha} 
\end{equation}     
Once $\gamma^{\omega}_x$ is balanced, the modulus of the third harmonic is given by:
\begin{equation}
|\gamma^{3\omega}|=\frac{1}{DR_\mathrm{S}+D'R_\mathrm{AC}}\frac{|\beta|R_\mathrm{AC}c'}{2}\label{eqgamma3w}
\end{equation}
At very low frequency ($\omega\sim0$), it is therefore possible to determine $R_\mathrm{AC}$ from the measurement of $\alpha(0)$ or of $|\gamma^{3\omega}|(0)$ 
considering that $|\beta(\omega=0)|=Re(\beta(\omega=0))=\frac{\Delta R_\mathrm{AC}}{R_\mathrm{AC}}$. At high frequency above the thermal cutoff frequency, $|\beta|\sim 0$. It results that $|\gamma^{3\omega}|\sim 0$. $R_\mathrm{AC}$ can be directly determined from $\alpha(+\infty)$.   

\subsection{Characterization of thermal properties of the SThM thermometer using the electro-thermal model}
\begin{figure}[h!]
\centering
\includegraphics[width=3.4in]{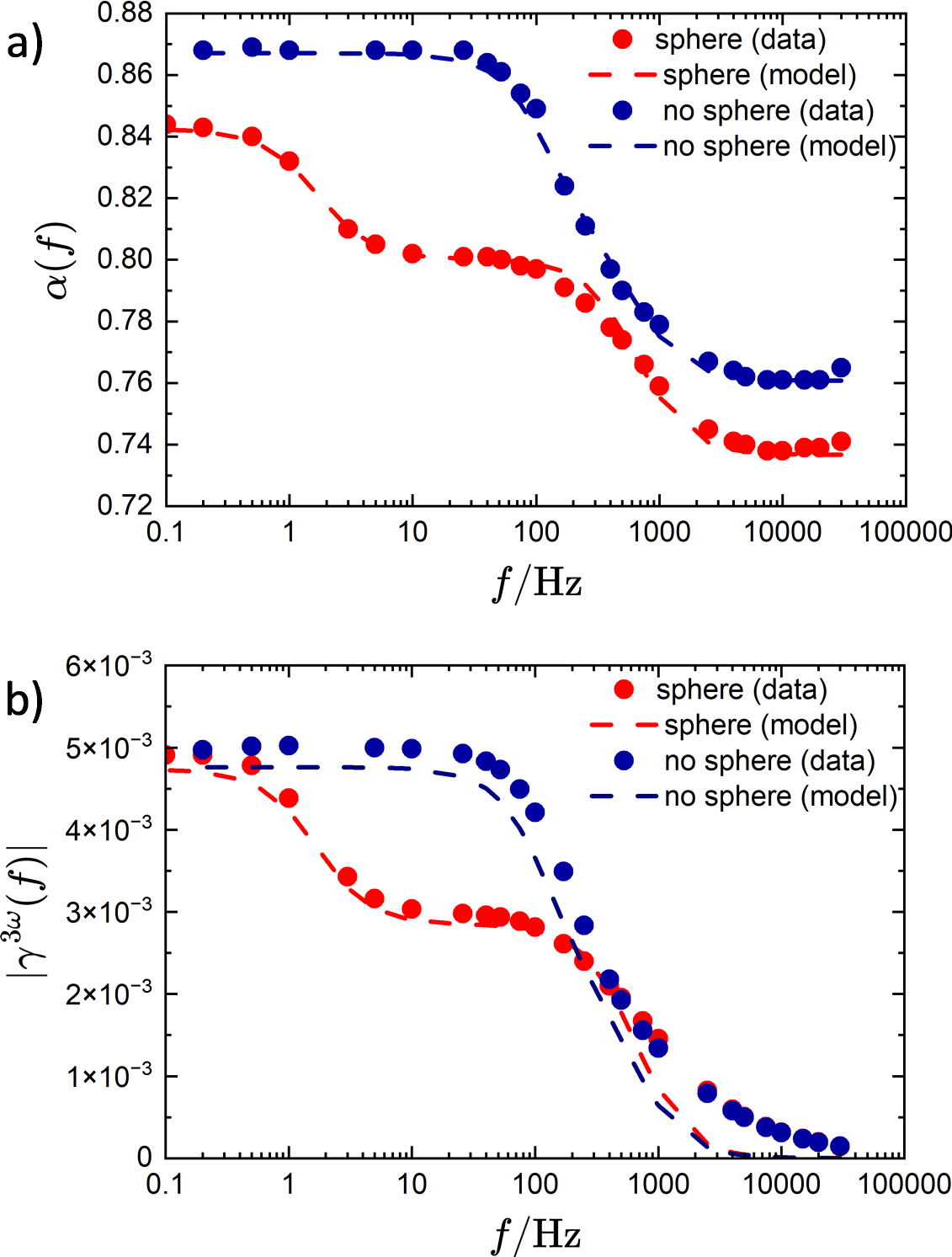}
\caption{$\alpha (f)$ (a) and $|\gamma^{3\omega}(f)|$ (b) measured as a function of frequency $f$ with a SThM probe equipped with (red circle) or not a glass microsphere (blue circle). Dashed lines represent adjustments obtained using equations (\ref{eqalpha}) and (\ref{eqgamma3w}).}\label{Frequence alpha gamma}
\end{figure}

The electro-thermal model is now used to characterize the dynamic thermal properties of the thermometer, equipped or not with a glass microsphere, from the frequency dependence of the electrical signals detected by the bridge. The bridge is biased with an AC voltage of frequency ranging from 0.1 Hz up to 15 kHz and of RMS amplitude $V_\mathrm{AC} = 0.125$ V. This leads to a small mean temperature increase of 7 K. For each frequency, the bridge is equilibrated and both the value of the potentiometer setting, $\alpha (f)$, and the relative bridge unbalance voltage, $|\gamma^{3\omega}(f)|$, are recorded. Figure \ref{Frequence alpha gamma}a) and b) report their frequency dependence both for the thermometer alone (blue) and for the same one equipped with a glass microsphere (red). The $\alpha (f)$ curve of the thermometer without the microsphere exhibits a single cutoff frequency. In contrast, two distinct characteristic frequencies clearly manifest in the frequency dependence measured with the microsphere glued on the SThM cantilever. The vertical shift between the two curves can be attributed to
a slight variation of the ambient temperature $T_0$ between the two measurements since $\alpha$ is indeed dependent on $R_0$. On the other hand, $|\gamma^{3\omega}(f)|$ is independent of the ambient temperature and depends only on the thermal properties of the probe. Its frequency dependence is similar to that of $\alpha (f)$ and exhibits same cutoff frequencies. As predicted by equation (\ref{eqgamma3w}), fig.~\ref{Frequence alpha gamma}b) shows that $|\gamma^{3\omega}(f)|$ tends towards zero above the highest cutoff frequency and towards a constant value at $f=0$, which is independent of the presence of the sphere. 

It is possible to satisfactorily adjust experimental curves using a thermal model of the complex conductance $G(\omega)$ based on the series of two elements of complex conductances $G_1+i\omega C_1$ and $G_2+i\omega C_2$, where $G_i$ and $C_i$ $(i=1:2)$ are thermal conductances and capacitances:
\begin{equation}
G(\omega)=\frac{(G_1+i\omega C_1)(G_2+i\omega C_2)}{(G_1+G_2)+i\omega(C_1+C_2)}
\end{equation}
The data adjustment using the model leads to the characteristic frequencies ($f_1=\frac{G_1}{4\pi C_1}$, $f_2=\frac{G_2}{4\pi C_2}$), equal to (645 Hz, 125 Hz) without microsphere and equal to (645 Hz, 1.7 Hz) with the microsphere glued onto the cantilever. The lowest cutoff frequency can be explained by the large thermal capacitance, $C_\mathrm{sphere}$, of the glass microsphere. The latter can be estimated from the specific heat capacity of borosilicate, $c_\mathrm{p}\simeq$ 830 $\mathrm{J\cdot kg^{-1}\cdot K^{-1}}$. For a 20 $\mu$m radius, one finds $C_\mathrm{sphere}\simeq 6\times10^{-8} \mathrm{J\cdot K^{-1}}$. From the cutoff frequency of 1.7 Hz, one would deduce a characteristic conductance $G_\mathrm{sphere}=4\pi C_\mathrm{sphere}\times 1.7\simeq 1.3~\mu \mathrm{W\cdot K^{-1}}$. This value is close to an estimation of the series conductance of the microsphere and of the epoxy (a disk of approximately 4 $\mu$m radius and 1 $\mu$m thickness) used to glue the microsphere onto the cantilever, which amounts to $\sim 2~\mu \mathrm{W\cdot K^{-1}}$. At higher frequencies, the cutoff observed around $645\,\mathrm{Hz}$, unchanged by the presence of the sphere, can be attributed to the thermal response of the cantilever itself. The cantilever can be modeled as a bimaterial beam with a rectangular cross section, composed of a gold layer of thickness $e_{\mathrm{gold}} = 150~\mathrm{nm}$ and a $\mathrm{Si_3N_4}$ layer of thickness $e_{\mathrm{Si_3N_4}} = 500~\mathrm{nm}$. Using reported thermal conductivities for gold and $\mathrm{Si_3N_4}$ ($100~\mathrm{W\,m^{-1}\,K^{-1}}$ and $2~\mathrm{W\,m^{-1}\,K^{-1}}$, respectively), one obtains a thermal cutoff frequency in the range of $400$-$600~\mathrm{Hz}$. This estimate is consistent with the experimentally measured cutoff.
  
\section{Application to the measurement of near-field radiative heat fluxes}
\begin{figure}[h!]
\centering
\includegraphics[width=3.4in]{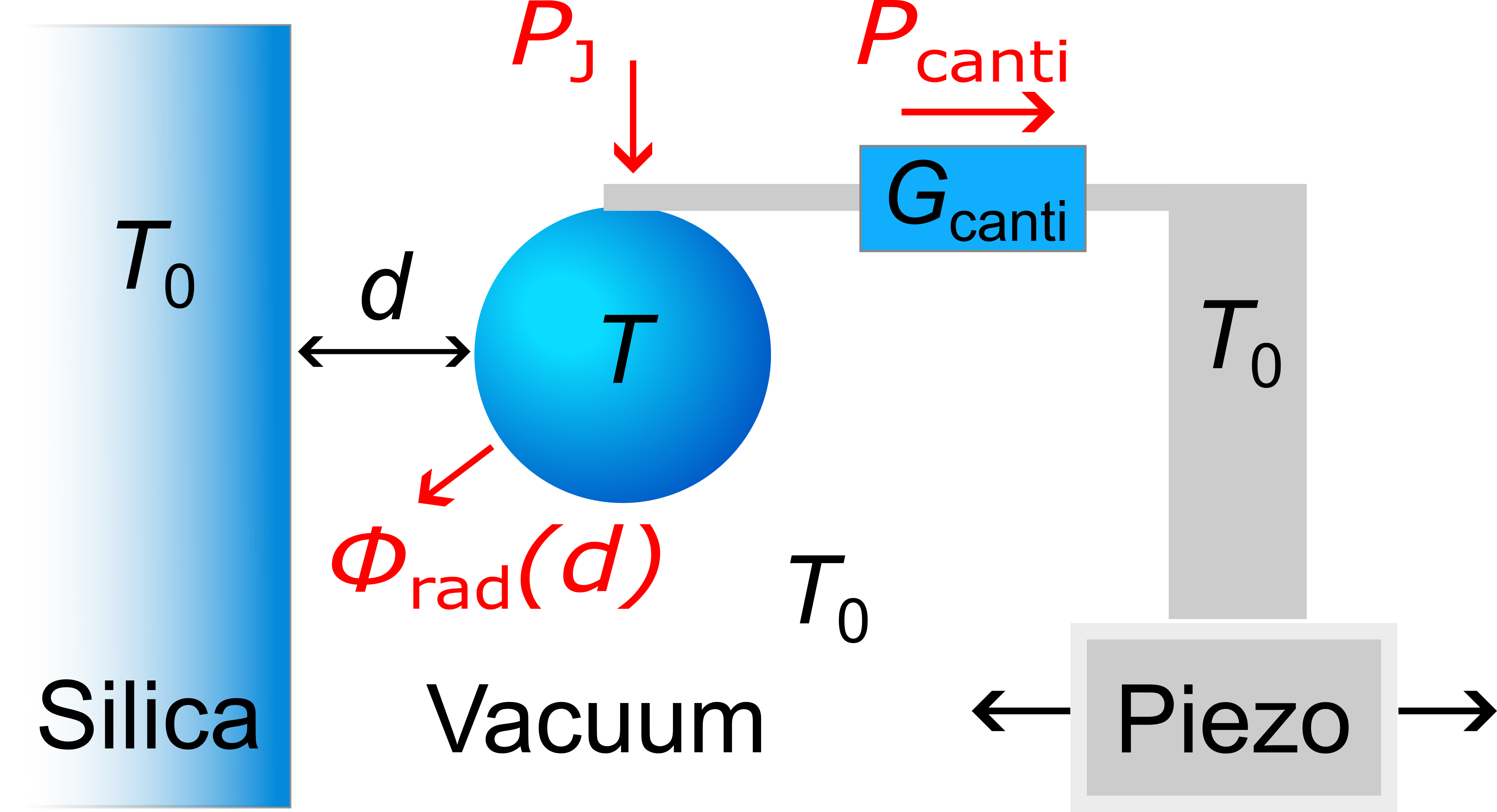}
\caption{Schematic of the measurement of the radiative heat flux exchanged between a silica microsphere and its environment. The thermal circuit shows that the thermal flux generated by Joule heating can flow by conduction through the cantilever of by radiation through the sphere towards environment and substrate.}\label{fig:figvic2} 
\end{figure}
\begin{figure}[h!]
\centering
\includegraphics[width=3.4in]{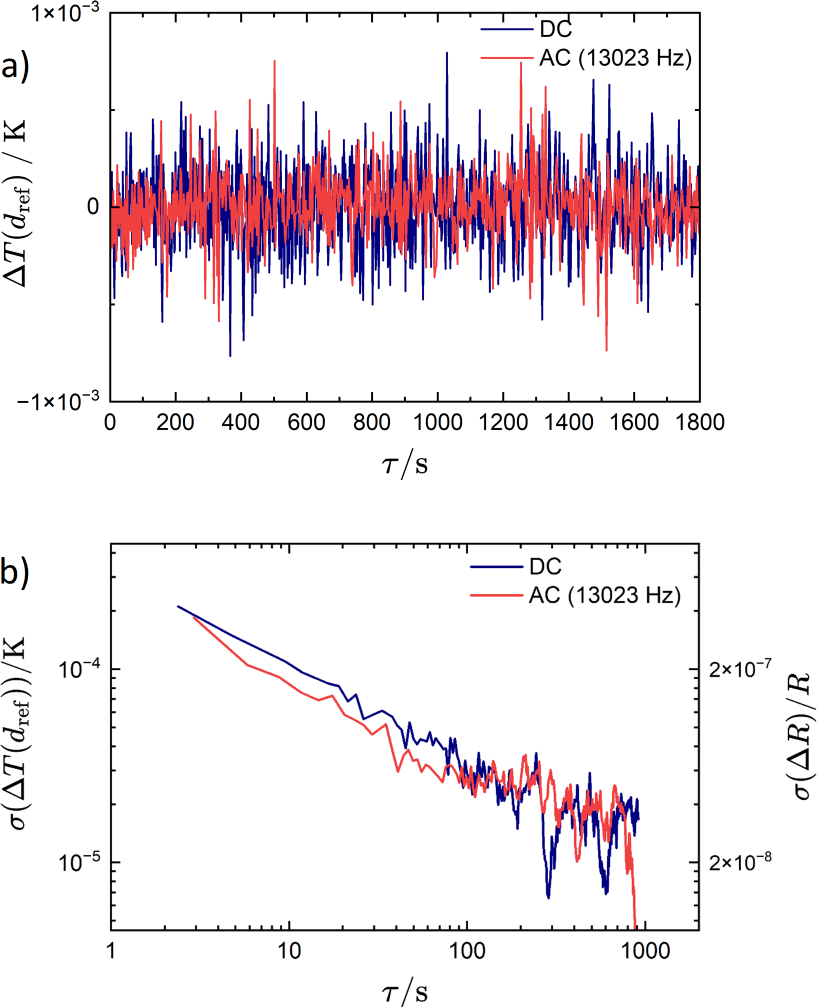}
\caption{a) Successive measurements of $\Delta T(d_\mathrm{ref})$ in DC (blue) and in AC (red) regime. b) Allan variance, $\sigma (\Delta T (d_\mathrm{ref}))$ (or $\sigma (\Delta R)/R$), as a function of time $\tau$ measured in DC regime (blue line) and in AC regime (red line) at 13 kHz frequency.}\label{Fluct Allan T}
\end{figure}

Fig.~\ref{fig:figvic2} describes a schematic of the experiment in vacuum, which consists in measuring the radiative heat flux between the 42.3 $\mu$m diameter glass microsphere glued on the cantilever described previously in fig.\ref{fig:figvic1} and its environment as a function of its distance, $d$, with a flat silica substrate. Let us remark that the microsphere is positioned at the end of the cantilever on the left side of the serpentine. Because the heat flux flowing towards the left side amounts to only one part in $10^3$ of the injected power (see the estimation of the radiative heat flux in the next paragraph), it results that the temperature of the microsphere is very close to that of the serpentine (typically to within one part in $10^3$). The distance between the sphere and the planar substrate is controlled by a 5 nm precision piezo-positioner. The temperature of the substrate and the environment are fixed at $T_0 = 23$ °C. The electrical resistance of the thermometer at $T_0$ is $R_0 = 283~\Omega $. The thermometer is connected to the Wheatstone bridge, which is either biased with a DC voltage or an AC voltage at a frequency of 13 kHz (well above the highest thermal frequency cutoff). Both measurements in DC and AC regime were performed using the same applied voltage on the platinum thermometer, $V_{\mathrm{AC/DC}} = 0.250~\mathrm{V}$, corresponding to a Joule power of $P_{\mathrm{J}} = 172~\mu\mathrm{W}$ leading to a constant temperature increase of the sphere of approximately $\Delta T = 32~^\circ\mathrm{C}$. The temperature is obtained from the resistance value using a linear relationship determined by an independent calibration of the thermometer. The temperature coefficient, $\lambda = (1.48\pm0.05)~\mathrm{K}\,\Omega^{-1}$, remains constant for moderate temperature variations (typically $\Delta T < 100^\circ\mathrm{C}$). The radiative flux $\phi_\mathrm{rad}(d)$ is obtained from the relationship:
\begin{equation}
\phi_\mathrm{rad} (d)= P_\mathrm{J}-G_\mathrm{canti}[T(d)-T_0],\label{equationphi} 
\end{equation}  
where $G_\mathrm{canti}$ is the cantilever conductance ($\sim 5~\mu \mathrm{WK^{-1}}$). It is determined in vacuum from the slope of the linear relationship between the injected power and the temperature increase ($\Delta P_\mathrm{J}^0/\lambda\Delta R^0$).
To circumvent time drift of $T_0$ and of $P_\mathrm{J}$, we proceed to a discrete spatial modulation of the cantilever. Two successive temperature measurements are successively performed at distance, $d$, and at a reference distance $d_\mathrm{ref}=4~\mu$m. After each cantilever displacement, the waiting delay before each measurement is 0.5 s. The measurement time is about 0.5 s. The total duration of one measurement cycle is 2 seconds. The quantity of interest is the flux variation given by:
\begin{equation}
\begin{split}
\Delta \phi_\mathrm{rad} (d)&=-G_\mathrm{canti}[T(d)-T(d_\mathrm{ref})] \\
&= - G_\mathrm{canti} \lambda [R(d) - R(d_\mathrm{ref})]
\end{split}
\label{eqn:dPhi}
\end{equation}
The type A uncertainty of $\Delta \phi_\mathrm{rad} (d)$ is therefore determined from that of the temperature difference $\Delta T(d)=T(d)-T(d_\mathrm{ref})$. To estimate this uncertainty over long time without being affected by the physical variation in temperature with distance, measurements of the temperature difference, $\Delta T(d_\mathrm{ref})= T(d)-T(d_\mathrm{ref})$, are repeated at the same distance $d=d_\mathrm{ref}$ using the same timing protocol. Fig. \ref{Fluct Allan T}a) reports successive measurements of $\Delta T(d_\mathrm{ref})$, which are within 1 mK, whatever they are performed in DC regime or in AC regime. The Allan deviation \cite{Witt2005}, $\sigma (\Delta T(d_\mathrm{ref}))/T$, is then calculated from these measurements as a function of the experiment time, $\tau$.   
It decreases following a $\tau^{-1/2}$ dependence, which proves that white noise dominates in the measurement of a temperature variation. Relative uncertainties of $4.5\times10^{-7}$ and $3\times10^{-8}$ can be achieved for 2 s and $1000~\mathrm{s}$ duration experiments, respectively. This corresponds to typically 200 $\mathrm{\mu K}$ (1 nW) and 15 $\mathrm{\mu K}$ (75 pW) temperature (heat power) uncertainties. Let us remark that similar uncertainties are achieved using either DC or AC measurements due to the efficiency of the spatial modulation protocol.         

One can then roughly estimate Type~B measurement uncertainties. For $\Delta T(d)$, it should consist of the uncertainty contributions of $\Delta R(d)= R(d)-R(d_\mathrm{ref})$ and of $\lambda$. Given the excellent accuracy of the resistance measurement, the second uncertainty contribution dominates and leads to a 
type B uncertainty, $u(\Delta T(d))=0.05\times \Delta T(d)$, which is no more than 2.5 mK for $\Delta T=50$ mK. Because $G_\mathrm{canti}$ is itself determined from $1/\lambda$, $\Delta \phi_\mathrm{rad}$ is not dependent on $\lambda$. We have therefore just to consider the reproducibility uncertainty of the cantilever thermal conductance, $G_\mathrm{canti}=(4.96\pm0.05)~\mu \mathrm{WK^{-1}}$, to estimate the Type B measurement uncertainty of $\Delta \phi(d)$. One finds $u(\Delta \phi_\mathrm{rad}(d))=0.01\times\Delta \phi_\mathrm{rad}(d)$, which amounts to 2 nW for $\Delta \phi_\mathrm{rad}=200~$ nW. Let us remark that relative variations of $\Delta T(d)$ and $\Delta \phi(d)$ (relatively to their maximum value) are independent of $\lambda$ and $G_\mathrm{canti}$ calibrations. Thus, the shape of the relative evolution of both quantities has lower measurement uncertainty and is therefore more robustly determined.

\begin{figure}[h!]
\centering
\includegraphics[width=3.4in]{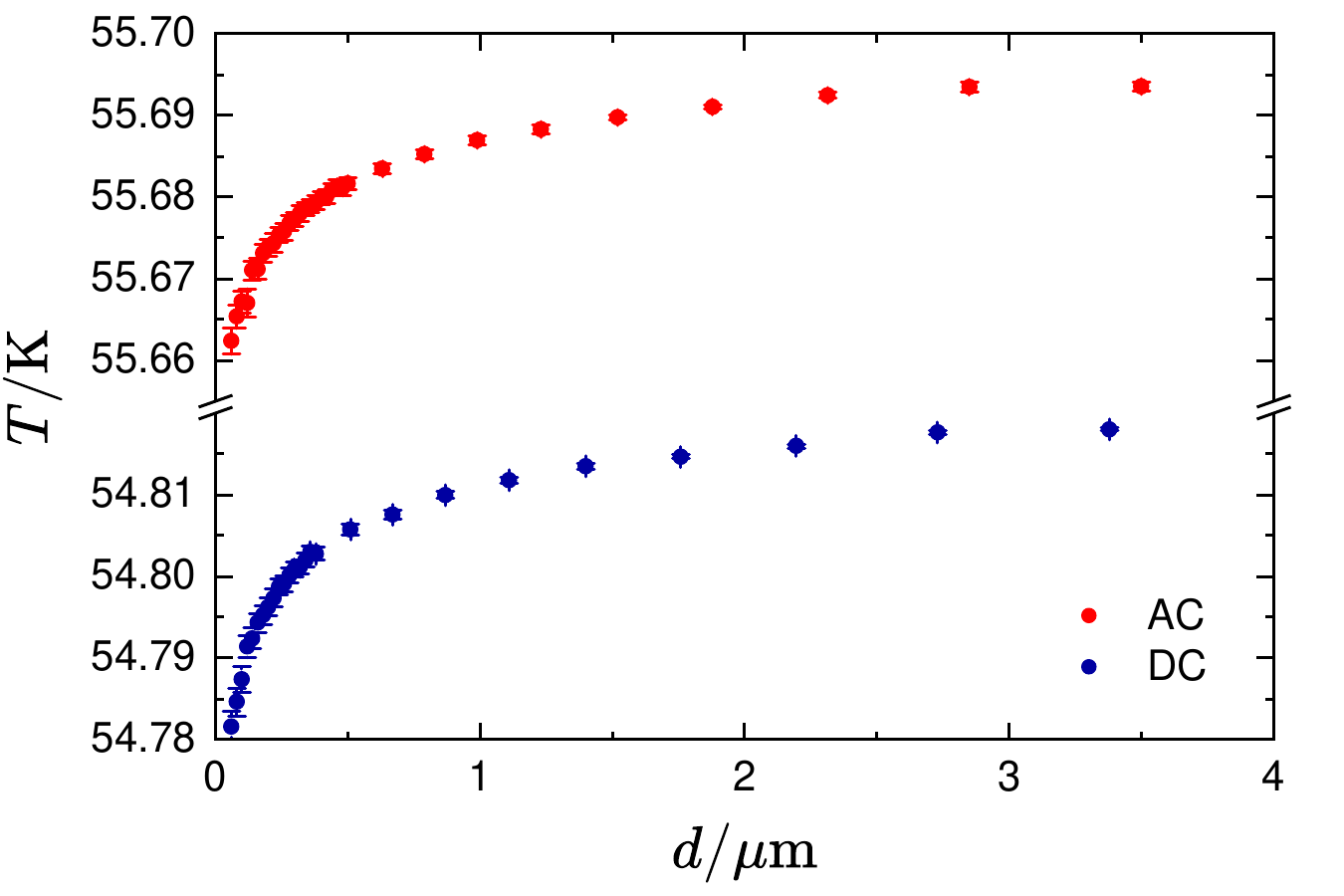}
\caption{Variation of the measured temperature as a function of the distance between the glass microsphere and the glass planar substrate. Measurements performed in AC at $13~\mathrm{kHz}$ (red circles) and in DC exhibit similar trends. The vertical offset between the two curves is attributed to a slight difference in the ambient temperature, $T_0$, between the two measurements. Errors bars corresponds to combined standard uncertainties.}\label{Approche T}
\end{figure}

Fig.~\ref{Approche T} shows approach curves of the temperature, $T$, measured in the last four micrometers using both DC and AC voltage biasing of the bridge. In order to cancel environment temperature drift, $T$ is obtained from the measurement of $\Delta T(d)$ and then using the relationship $T(d)=\Delta T(d) + T(d_\mathrm{ref},0)$, where $d_\mathrm{ref}=4~\mu$m and $T(d_\mathrm{ref},0)$ is the temperature at the reference distance at the beginning of the experiment. Error bars represent combined standard uncertainties (coverage factor k=1), which include both Type A and Type B uncertainties previously described. Apart a small temperature shift of the initial ambient temperature, the two curves are very similar and show a temperature decrease as the distance decreases, which is explained in the following by an increase of the near-field radiative heat flux.
\begin{figure}[h!]
\centering
\includegraphics[width=3.4in]{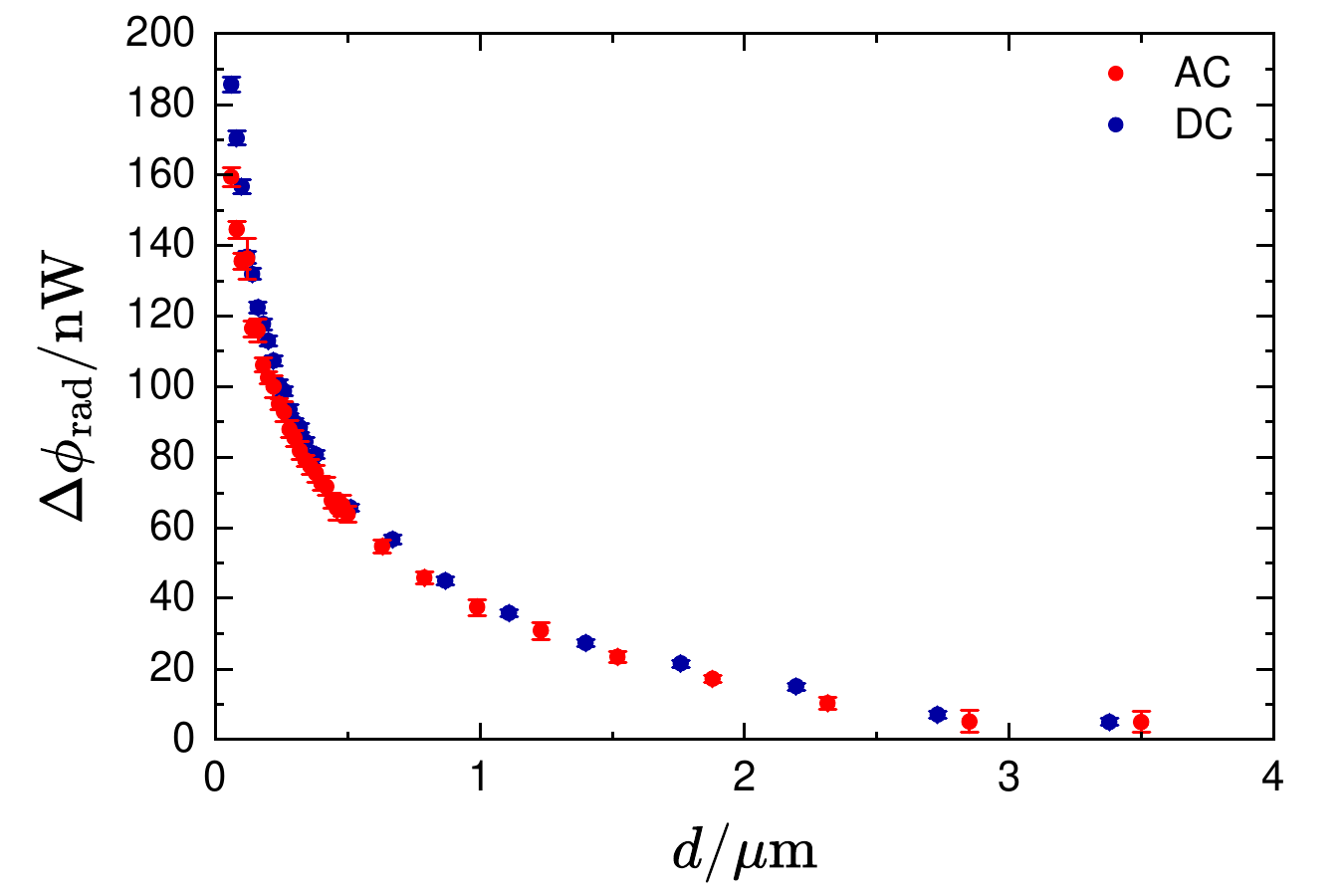}
\caption{Radiative flux variation, $\Delta \phi_\mathrm{rad}$ measured between distance, $d$, and the reference distance $d_\mathrm{ref} = 4~\mu m$. The radiative flux is determined from data of figure \ref{Approche T} and equation (\ref{eqn:dPhi}). Errors bars correspond to combined standard uncertainties.}\label{Approche Phi}
\end{figure}

The radiative heat flux variation, $\Delta \phi_\mathrm{rad}$, calculated using equation (\ref{eqn:dPhi}) is reported in fig. \ref{Approche Phi}. Both DC and AC measurements reveal an increasing radiative heat flux in the near-field regime of about 180 nanowatts, corresponding to only about one part in $10^3$ of the total injected power. This can be explained by the coupling of phonon-polaritons at surfaces of the substrate and of the microsphere\cite{Mulet2002,Joulain2005}. With the support of theoretical and numerical tools, this is demonstrated and discussed in detail in work\cite{Guillemot2025}, which reports on radiative heat flux measurements that we performed using a first and simpler Wheatstone bridge prototype\cite{GuillemotPhd2025}.    
\section{Conclusion}
We have developed a new resistance bridge optimized for high-precision resistance measurements of thermometers in scanning thermal microscopes. 
Based on a Wheatstone bridge architecture allowing approximate four-wire resistance definition, it is able to operate not only in direct current but also 
in alternating regime at frequencies up to a few tenths of kHz. We have demonstrated that standard resistances in the range from 100 $\Omega$ up to 1000 $\Omega$ 
can be accurately measured, both in DC and in AC current regime, with relative type B uncertainties below 50 parts in $10^{6}$. The relative experimental standard 
deviation can be as low as one part in $10^8$ for one second measurement. The measuring instrument was tested with SThM resistance thermometers. Using a simple 
electro-thermal model, we explain the frequency dependence of measurements from the thermal time constants of the thermometer. This analysis demonstrates that 
the temperature expression derived from the bridge parameters is identical for both DC measurements and AC measurements performed at frequencies above the highest 
thermal cutoff frequency. Finally, we exploit the new instrument to measure radiative heat transfers in the near-field regime between a glass substrate at ambient 
temperature and a heated glass microsphere, coming closer, which is glued onto the end of a cantilever close to the thermometer. More precisely, the instrument 
is used to heat the microsphere by injecting a defined Joule power in the thermometer and measure its temperature as a function of the distance to the substrate. 
Allan deviation analysis shows that temperature (heat power) variation can be measured with relative uncertainties as low as 200 $\mathrm{\mu K}$ (1 nW) uncertainties 
for measurement of two seconds. This performance allows us to observe the increase of radiative heat transfer in the near-field regime caused by the coupling of 
phonon-polaritons at surfaces of the substrate and of the microsphere. An experimental platform equipped with two nano-positioned SThM probes, each one connected 
to is own measuring instrument, has been recently developed\cite{GuillemotPhd2025}. It successfully allowed the study of radiative heat transfers in systems 
consisting of several bodies at different temperatures \cite{Guillemot2026}.  
\section{APPENDIX}
\begin{figure}[h]
\centering
\includegraphics[width=3.3in]{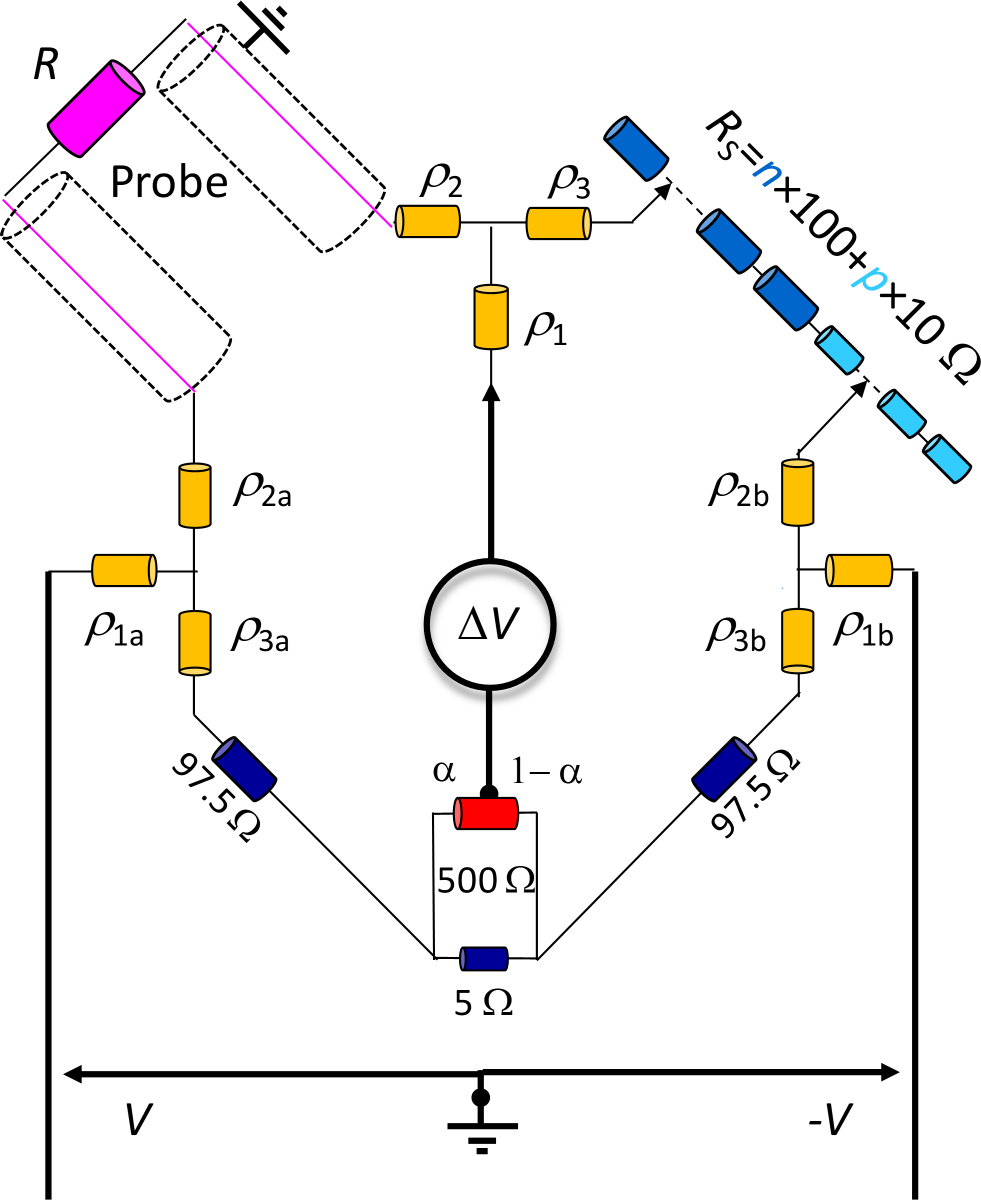}
\caption{Equivalent circuit of the resistance bridge considering triangle-star transformations.}\label{figEWheatstone}
\end{figure}

\textbf{APPENDIX A: Equivalent circuit of the resistance bridge}\\ \\
\label{annexe:EquivalentWheatstone}
\normalsize
Fig.\ref{figEWheatstone} shows the schematic of the equivalent circuit of the resistance bridge. After applying triangle-star transformations (Kennely transformation), 
the Thomson bridge\cite{Trapon1997} reduces to a Wheatstone bridge. 
The relationships between the $\rho$ and the $r$ resistances are given by:
\tiny
\begin{eqnarray}
\rho_1&=&\frac{(100+r_2)(100+r_3)}{r_1+r_2+r_3+200}\simeq 50~\Omega\\
\rho_2&=&\frac{r_1(r_2+100)}{r_1+r_2+r_3+200}\simeq \frac{r_1}{2}[1-\frac{(r_1+r_3-r_2)}{200}]\\
\rho_3&=&\frac{r_1(r_3+100)}{r_1+r_2+r_3+200}\simeq \frac{r_1}{2}[1-\frac{(r_1+r_2-r_3)}{200}]\\ \\
\rho_\mathrm{1a}&=&\frac{10(R_\mathrm{S}/10+r_\mathrm{2a})}{10+R_\mathrm{S}/10+r_\mathrm{1a}+r_\mathrm{2a}}\simeq \frac{10R_\mathrm{S}}{100+R_\mathrm{S}}~\Omega\\
\rho_\mathrm{2a}&=&\frac{r_\mathrm{1a}(R_\mathrm{S}/10+r_\mathrm{2a})}{10+R_\mathrm{S}/10+r_\mathrm{1a}+r_\mathrm{2a}}\simeq r_\mathrm{1a}\frac{R_\mathrm{S}}{100+R_\mathrm{S}}[1-10\frac{(r_\mathrm{1a}-100r_\mathrm{2a}/R_\mathrm{S})}{100+R_\mathrm{S}}]\\
\rho_\mathrm{3a}&=&\frac{10r_\mathrm{1a})}{10+R_\mathrm{S}/10+r_\mathrm{1a}+r_\mathrm{2a}}\simeq r_\mathrm{1a}\frac{100}{100+R_\mathrm{S}}[1-10\frac{(r_\mathrm{1a}+r_\mathrm{2a})}{100+R_\mathrm{S}}]\\\\
\rho_\mathrm{1b}&=&\frac{10(R_\mathrm{S}/10+r_\mathrm{2b})}{10+R_\mathrm{S}/10+r_\mathrm{1b}+r_\mathrm{2b}}\simeq \frac{10R_\mathrm{S}}{100+R_\mathrm{S}}~\Omega\\
\rho_\mathrm{2b}&=&\frac{r_\mathrm{1b}(R_\mathrm{S}/10+r_\mathrm{2b})}{10+R_\mathrm{S}/10+r_\mathrm{1b}+r_\mathrm{2b}}\simeq r_\mathrm{1b}\frac{R_\mathrm{S}}{100+R_\mathrm{S}}[1-10\frac{(r_\mathrm{1b}-100r_\mathrm{2b}/R_\mathrm{S})}{100+R_\mathrm{S}}]\\
\rho_\mathrm{3b}&=&\frac{10r_\mathrm{1b})}{10+R_\mathrm{S}/10+r_\mathrm{1b}+r_\mathrm{2b}}\simeq r_\mathrm{1b}\frac{100}{100+R_\mathrm{S}}[1-10\frac{(r_\mathrm{1b}+r_\mathrm{2b})}{100+R_\mathrm{S}}]
\end{eqnarray}
\normalsize
The resistance, $r_1$, of the current wire connecting the thermal probe is therefore equally shared between $R$ and $R_\mathrm{S}$ since $\rho_2$ is equal to $\rho_3$ to within 
a relative correction of $\mathcal{O}(r/200)$. In the same way, $\frac{\rho_{2a}}{\rho_{3a}}$ and $\frac{\rho_{2b}}{\rho_{3b}}$ are equal 
to the resistance ratio $\frac{R_\mathrm{S}}{100}$ to within a relative correction of $\mathcal{O}(10r/R_\mathrm{S})$. Kelvin arms therefore allows to reduce 
strongly the sensitivity of the bridge to wire resistances of the thermal probe as well as to contact resistances defining the reference standard resistance, $R_\mathrm{S}$. 
More precisely, one can calculate the relative error in the estimation of the resistance, $\Delta R/R_\mathrm{S}$, resulting from the wire and contact resistances. Keeping the lowest order, one obtains: 
\begin{equation}
\tiny
\Delta R/R_\mathrm{S}\simeq \frac{10(r_{1a}r_{2a}-r_{1b}r_{2b})}{R_\mathrm{S}(100+R_\mathrm{S})}+\frac{r_1(r_2-r_3)}{200R_\mathrm{S}}+\frac{(r_{1b}-r_{1a})}{100+R_\mathrm{S}}(\frac{r_1}{2R_\mathrm{S}}+\frac{r_{1b}}{100+R_\mathrm{S}}) 
\label{eqDR}
\end{equation}
\normalsize
Equation (\ref{eqDR}) shows that the implementation of Kelvin arms cancels first order terms  $\mathcal{O}(r/R_\mathrm{S})$ in the expression of $\Delta R/R_\mathrm{S}$. 
Only second order corrections, $\mathcal{O}(r^2/100R_\mathrm{S})$, $\mathcal{O}(r^2/(100+R_\mathrm{S})R_\mathrm{S})$ and $\mathcal{O}(r^2/(100+R_\mathrm{S})^2)$ 
can shift the measured resistance value of the thermal probe ensuring bridge balance. Note that the first term gives the main contribution. A 2 $\Omega$ increase of 
each one of the probe wires would result in a deviation of $\Delta R/R_\mathrm{S}$ by $2.1\times10^{-3}$, $1.67\times10^{-4}$ and $5.45\times10^{-5}$ for resistance 
values of 100 $\Omega$, 500 $\Omega$ and 1000 $\Omega$, respectively.  \\ \\
\textbf{APPENDIX B: In-phase and out-phase expressions for $\gamma$}\\ \\
\label{annexe:gammaexpression}
From the expression (\ref{eqR}) of the resistance $R$, one can obtain the in-phase and out-phase components of $\gamma$ at frequency $\omega$ and $3\omega$ using the first approximation in $\beta$:
\tiny
\begin{eqnarray}
\gamma^{\omega}_x &=&\frac{1}{DR_\mathrm{S}+D'R_\mathrm{AC}}[(R_\mathrm{S}c-R_\mathrm{AC}c')-\frac{Re(\beta) R_\mathrm{AC}}{2}(c'+\frac{D'(R_\mathrm{S}c-R_\mathrm{AC}c')}{DR_\mathrm{S}+D'R_\mathrm{AC}})]\\
\gamma^{\omega}_y &=&\frac{1}{DR_\mathrm{S}+D'R_\mathrm{AC}}[\frac{Im(\beta) R_\mathrm{AC}}{2}(c'+\frac{D'(R_\mathrm{S}c-R_\mathrm{AC}c')}{DR_\mathrm{S}+D'R_\mathrm{AC}})]\\
\gamma^{3\omega}_x &=&\frac{1}{DR_\mathrm{S}+D'R_\mathrm{AC}}[-\frac{Re(\beta) R_\mathrm{AC}}{2}(c'+\frac{D'(R_\mathrm{S}c-R_\mathrm{AC}c')}{DR_\mathrm{S}+D'R_\mathrm{AC}})]\\
\gamma^{3\omega}_y &=&\frac{1}{DR_\mathrm{S}+D'R_\mathrm{AC}}[\frac{Im(\beta) R_\mathrm{AC}}{2}(c'+\frac{D'(R_\mathrm{S}c-R_\mathrm{AC}c')}{DR_\mathrm{S}+D'R_\mathrm{AC}})]
\end{eqnarray}
\normalsize
Close to in-phase equilibrium at pulsation $\omega$ (i.e. $\gamma^{\omega}_x\simeq 0$), $(R_\mathrm{S}c-R_\mathrm{AC}c')\simeq \frac{Re(\beta) R_\mathrm{AC}c'}{2}$. 
It results that the last term in the equations is at the second order in $\beta$.\\ \\

\large\textbf{Acknowledgment}\\
\normalsize
The authors thanks Nolwenn Fleurence for fruitful discussions and critical reading of the article. This work was supported by the Agence Nationale de la Recherche (NBODHEAT Project No. ANR-21-CE30-0030). \\
\providecommand{\noopsort}[1]{}\providecommand{\singleletter}[1]{#1}%
\end{document}